\documentclass[aip,jcp,amsmath,amssymb,reprint]{revtex4-1}
\pdfoutput=1
\usepackage[utf8]{inputenc}
\usepackage[T1]{fontenc}
\usepackage{ulem}
\usepackage{gensymb}
\usepackage{xcolor}
\usepackage{graphicx}
\usepackage{amsthm}
\usepackage{dcolumn}
\usepackage{epsfig}
\usepackage{bm}
\usepackage{datetime}
\usepackage[title]{appendix}
\usepackage{booktabs}
\usepackage{ulem}

\begin{document}

\title{Practical guide to replica exchange transition interface sampling and forward flux sampling}
\author{Steven W. Hall}
\affiliation{Department of Chemical Engineering and Materials Science, University of Minnesota, Minneapolis, MN 55455, USA}
\author{Grisell D{\'i}az Leines}
\affiliation{Yusuf Hamied Department of Chemistry, University of Cambridge, Cambridgeshire CB2 1EW, United Kingdom}
\author{Sapna Sarupria}
\email{sarupria@umn.edu}
\affiliation{Department of Chemistry, University of Minnesota, Minneapolis, MN 55455, USA}
\affiliation{Department of Chemical and Biomolecular Engineering, Clemson University, Clemson, SC 29634, USA}
\author{Jutta Rogal}
\email{jutta.rogal@nyu.edu}
\affiliation{Department of Chemistry, New York University, New York, NY 10003, USA}
\affiliation{Fachbereich Physik, Freie Universit{\"a}t Berlin, 14195 Berlin, Germany}
\begin{abstract}
Path sampling approaches have become invaluable tools to explore the mechanisms and dynamics of so-called {\it rare events} that are characterized by transitions between metastable states separated by sizeable free energy barriers.  Their practical application, in particular to ever more complex molecular systems, is, however, not entirely trivial. Focusing on replica exchange transition interface sampling (RETIS) and forward flux sampling (FFS), we discuss a range of analysis tools that can be used to assess the quality and convergence of such simulations which is crucial to obtain reliable results.  The basic ideas of a step-wise evaluation are exemplified for the study of nucleation in several systems with different complexity, providing a general guide for the critical assessment of RETIS and FFS simulations.   
\end{abstract}

\maketitle

\section{Introduction}
 The standard approach to study the microscopic dynamical behaviour of classical particle systems is molecular dynamics (MD) simulations.
 But even with rapidly growing computational resources, the timescale accessible in MD simulations remains limited. Many processes of interest, including nucleation-driven phase transformations, chemical reactions, and conformational changes in biomolecular systems, take place on long timescales inaccessible to straightforward MD simulations. Such processes are usually characterized by transitions between metastable states separated by significant free energy barriers. This results in a separation of timescales between the short-time dynamics within the states and the infrequent transitions between the states, termed {\it rare events}.

A number of computational approaches has been developed over the years to facilitate the sampling of rare events. One class of methods drives the system along a predefined reaction coordinate (RC), given by a set of a few collective variables (CVs), to efficiently probe the free energy landscape and provide information on the mechanism.\footnote{RC here refers to the combination of CVs along which the transition occurs. If the dynamics of a system along the RC can be described by overdamped Langevin dynamics, the committor constitutes the ideal RC. Any combination of CVs that exhibits a high correlation with the committor can, therefore, be considered as a reasonable approximation to the RC.} 
Examples of such methods include metadynamics,~\cite{Laio2002,Laio05,Laio08,Barducci08} (driven) adiabatic free energy dynamics,~\cite{Rosso02a,Rosso02b,Abrams2008} and temperature-accelerated MD~\cite{Maragliano06}. These methods do not yield direct dynamical information of the process. In case of high-dimensional systems, the choice of suitable CVs to construct the RC can be very challenging but is crucial as the results inherently depend on this choice.
The second class of rare event methods uses simulations with unbiased dynamics to sample transition pathways while preserving dynamical properties. Methods that fall into this category include transition path sampling (TPS),~\cite{Dellago98a,Dellago98b,Bolhuis02,Dellago2002} transition interface sampling (TIS),~\cite{Bolhuis:03:JCP,Bolhuis:05:JComputPhys} forward flux sampling (FFS),~\cite{Allen2005,Allen2006a,Allen2006b,Allen2009} weighted ensemble,~\cite{Kim:96:BPJ,Zuckerman:10:JCP,Chong:17:AnnRev} and milestoning~\cite{Elber:04:JCP,Elber:15:JCP,Elber:20:AnnRev}. These methods avoid an a priori definition of an RC by generating an ensemble of unbiased, dynamical trajectories between metastable states. The path ensemble contains information about the mechanisms as well as thermodynamic and kinetic properties of the transition. Over the years, many extensions to these approaches have been introduced and we refer to Refs.~\onlinecite{Bolhuis2010,Peters2010,VanErp2012,HajiAkbari:20:JCP,Bolhuis:21:ATS} for extensive reviews of  trajectory-based sampling approaches. 

Two path sampling methods with the same theoretical underpinnings, replica exchange transition interface sampling (RETIS)\cite{VanErp2007,Bolhuis2008} and FFS, have become particularly popular and efficient for the calculation of rate constants. These methods have been used to study a wide variety of processes such as protein folding,~\cite{Escobedo:10:JCP,Escobedo:09:JPhysCondMat,Bolhuis:08:BPJ} nucleation,~\cite{DeFever:17:JCP,DiazLeines2017,DiazLeines2018,Rogal:20:JCPa,Rogal:20:JCPb,AZP:18:JCP,Debenedetti:15:PNAS} and chemical reactions~\cite{vanErp:18:PNAS,Tidor:19:JACS}. While the theoretical foundations of these approaches are well-established and exact, their practical application can be challenging. In recent years, significant effort has been allocated to the development of software packages that implement these techniques and make them accessible to a wide range of users,~\cite{Lervik2017,Riccardi2020,Swenson2019a,Swenson2019b} complemented by helpful discussions about conceptual pitfalls~\cite{Bolhuis11,Bolhuis2015}. Still, for new users, practical issues frequently arise that are not easily detected or solved. One of the key issues is sufficient decorrelation in the path ensembles which, with the exception of simple model systems, is neither trivial to accomplish nor to evaluate. This challenge is further intensified because, unlike simple model systems where virtually infinite sampling is achievable, path sampling approaches applied to complex (and more realistic) systems are computationally expensive and require an efficient exploration of trajectory space. Motivated by this and our own experiences in applying these methods to complex systems, we provide a practical guide to evaluate and improve the sampling obtained in RETIS and FFS simulations with the aim to support new and less experienced users in obtaining reliable results.

As examples to illustrate the range of sampling behaviors that might be observed in RETIS and FFS simulations, we consider nucleation in several systems with different complexity.
For RETIS, we discuss expected behaviors at convergence as well as various analysis steps that can be used to detect problematic behavior within the limit of finite sampling.
For FFS, we evaluate the convergence at different stages of the simulations and introduce several ideas to detect possible sampling issues on-the-fly.
Since each new system exhibits a different behavior, the solutions to the sampling problems for the presented case studies will not be one-to-one transferable. However, the overall procedure to detect and analyze suspicious behavior that can point towards poor sampling is generally applicable. Such  recipes are particularly useful when studying new and ever more complex systems to quickly detect sampling issues and improve the sampling efficiency while keeping the computational cost in a manageable range.

The article is organized as follows. In the Sec.~\ref{sec:background}, we briefly recapitulate the basic principles of RETIS and FFS. In Sec.~\ref{sec:practicalguide} we introduce our example systems and present the practical step-by-step guide to assess the quality of RETIS and FFS simulations and address potential problems. We conclude with a short discussion of the critical quantities and parameters to be considered in the two approaches. 

%-------------------------- Theoretcial Background --------------------

\section{Theoretical Background}
\label{sec:background}

The central idea of both RETIS and FFS is to sample an ensemble of characteristic dynamical pathways between stable states and obtain atomistic insight into the transition mechanism. The path ensemble can, among other things, be used to compute dynamical properties of the system, such as rate constants. In RETIS and FFS, trajectories are sampled  using the CV $\lambda$ that discriminates between the stable states. However, both approaches do not require that $\lambda$ closely matches the RC, which is advantageous for the investigation of complex mechanisms with nontrivial RCs. The stable states are defined in terms of $\lambda(x)$ as regions of the phase space such that $\lambda(x)\leq \lambda_A$ in state $A$ and $\lambda(x)\geq \lambda_B$ in state $B$, where $x$ is a point in phase space, $x=\{r^N,p^N\}$, for a system with $N$ atoms. Both methods compute the rate constant using the effective positive flux definition, introduced by van Erp et al.~\cite{Bolhuis:05:JComputPhys} within the TIS framework:
\begin{equation}
\label{eq:rateconstant}
k_{AB}=\Phi_0 P_{A}(\lambda_n|\lambda_0) \quad .
\end{equation}
Here, $\lambda_0=\lambda_A$ and $\lambda_n=\lambda_B$ are the stable state boundaries, $\Phi_0$ is the positive flux of trajectories leaving state $A$ and crossing the first interface $\lambda_0$, and $P_{A}(\lambda_n|\lambda_0)$ is the crossing probability, i.e., the conditional probability that a trajectory coming from state $A$ reaches state $B$ before returning to $A$.  Both RETIS and FFS achieve an efficient computation of the rate constant by dividing the phase space between the stable states into a series of non-intersecting interfaces, $\{\lambda_1(x),...,\lambda_n(x)\}$,   and calculating the overall crossing probability $P_{A}(\lambda_n|\lambda_0)$ as the product of individual probabilities $P_{A}(\lambda_{i+1}|\lambda_i)$ between interfaces $i$ and $i+1$,
\begin{equation}
\label{eq:crossingp}
 P_{A}(\lambda_n|\lambda_0)=\prod_{i=0}^{n-1} P_{A}(\lambda_{i+1}|\lambda_i) \quad .
\end{equation}
Although both methods are based on an interface sampling scheme, there are several differences that result in respective strengths and limitations of each algorithm. RETIS samples the path ensemble by performing 
Monte Carlo (MC) sampling in trajectory space. 
The sampling assumes an equilibrium phase space distribution, which limits its applicability to Markovian processes. However, the microscopic reversibility required in RETIS guarantees a correct relaxation of the paths to an unbiased path ensemble and the algorithm can handle any type of reversible dynamics (including deterministic). In FFS, trajectories are sampled by `shooting' partial trajectories between interfaces, but only in the forward direction in time. Consequently, FFS does not require microscopic reversibility and does not assume an equilibrium density distribution, which allows for the study of non-equilibrium processes and makes the algorithm conceptually and practically simpler. Yet, the restriction to forward sampling suffers from a strong dependency on the initial stages of the simulation. If, for example, the initial pathways that cross the first interface are not representative of the equilibrium ensemble, a strong bias is introduced and relaxation of the pathways becomes very difficult. Having forward relaxation of the paths also implies that the convergence of FFS simulations is more sensitive to the choice of the CV than RETIS, because poorly defined interfaces $\lambda_i$ can result in non-representative distributions of shooting points at the  interfaces. 
In FFS, stochastic dynamics is required  to achieve ergodic sampling. The placement of interfaces serves slightly different purposes for each method. The optimization of interface placement in RETIS is mainly used to reduce the variance in the estimation of the crossing probabilities and correspondingly, to enhance the efficiency of the sampling. In FFS, the placement of interfaces is optimized as well to improve efficiency, but also to reduce the statistical error in the estimation of effective positive flux of trajectories for later interfaces, and thus the estimation of the rate constant. Next, we briefly review  the basic ideas of the  RETIS and FFS algorithms and the technical advances that have been introduced to improve sampling quality and efficiency.

%--------------------- TIS ----------------------------------------------

\subsection{Replica exchange transition interface sampling}

Conventional TPS performs MC sampling in trajectory space to generate an ensemble of dynamical trajectories connecting two (meta-)stable states. This ensemble is the transition path ensemble (TPE). A {\it path} is a discretized trajectory, $\mathbf{x}^{(L)} = \{x_1,\dots,x_L\}$, that consists of $L$ successive phase space points $x$ separated by a time interval $\Delta t$. The configuration space is divided into two stable state regions that are defined by 
$\lambda(x)$ and, in analogy to configurational MC, the MC sampling of transition paths follows two basic steps:
({\it i}) a trial path is generated from an existing path by a MC move;
({\it ii}) the new trial path is accepted or rejected according to the corresponding detailed balance criterion of the TPE.
Although there are several algorithms to generate new trial paths, the {\it shooting} algorithm is popular for its simplicity and efficiency. In the original implementation of the shooting algorithm,~\cite{Dellago2002} a phase space point from the current path is randomly selected and the momenta and/or positions are slightly perturbed. A new trajectory is created from the modified configuration by integrating the equations of motion both forward and backward in time.  If the new path connects the two stable states it is accepted within the detailed balance criterion, otherwise the old path is kept and recounted in the TPE. This algorithm is one of the two-way shooting methods since trajectories are generated both forward and backward in time from the shooting point.
The shooting move is repeated in the next MC step as the sampling of the TPE progresses.
Ultimately, TPS aims to achieve  an {\it efficient} and {\it ergodic} sampling of crossing pathways in the barrier region for energy landscapes with a variety of complex features. 
Depending on the characteristics of the energy landscape, the efficiency can significantly vary with the selected MC moves, resulting in the proposal of many modifications to the original shooting algorithm. These either promote the selection of configurations near the barrier region, thus increasing the chance of creating accepted new trial paths, or enhance the decorrelation of successive paths to improve ergodicity.
One example is the introduction of flexible path lengths to TPS.  By allowing paths with flexible duration that are stopped as soon they reach the stable states, the probability of accepted trial shots is increased, thereby saving a considerable amount of computation time.~\cite{Bolhuis:05:JComputPhys,Bolhuis2010} With regard to decorrelation,  the stochastic behavior of molecular systems usually ensures that new paths differ sufficiently from previous ones and properly sample the trajectory space. However, for diffusive barriers that are broad or rough, large divergence of the paths often results in frequent unsuccessful trial paths and, consequently, a very low acceptance. For these cases, precision shooting~\cite{Gruenwald2008} was introduced to create new trial trajectories arbitrarily close to the old ones and  increase the acceptance and efficiency of the algorithm considerably. Other methods, like spring shooting~\cite{Brotzakis2016} or aimless shooting~\cite{Peters06,Mullen2015}, enhance the success probability by biasing the shooting points towards the barrier region.  In one-way shooting,~\cite{Bolhuis2003} only a part of the trajectory is replaced in each move, which is beneficial when sampling very long trajectories.  
Other shooting algorithms, like web-throwing and stone-skipping,~\cite{Riccardi2017} focus on achieving fast decorrelation of trajectories by generating new paths from subsequent short and highly accepted sub-trajectories. A recent overview over the various shooting algorithms and available TPS and TIS software is given in Ref.~\onlinecite{Bolhuis:21:ATS}.

TIS~\cite{Moroni2004,Bolhuis:05:JComputPhys,VanErp2007} was introduced to provide a more efficient way to calculate rate constants compared to TPS. In TIS, 
each interface $\lambda_i$ is associated with a separate path ensemble, where any trajectory that crosses the interface is included in the ensemble, even if it returns to its initial stable state. The MC sampling for each interface ensemble proceeds in the same fashion as outlined above for TPS. 
The rate constant is then obtained from the crossing probabilities of trajectories through each interface between the stable states, as given by Eqs.~\eqref{eq:rateconstant} and~\eqref{eq:crossingp}. 

In addition to shooting moves, exchange moves between the interface ensembles, introduced in the RETIS algorithm,~\cite{VanErp2007,Bolhuis2008}  considerably increase the efficiency and enhance the ergodicity of the sampling. 
Trajectories can be exchanged between interface ensembles if they mutually cross the corresponding interfaces.  Usually, swaps are attempted between neighboring interfaces, since these are placed such that  the ensembles sufficiently overlap. Overlap between ensembles $i$ and $i+1$ in this case refers to the crossing probabilities, $P_{A}(\lambda_{i+1}|\lambda_i)$.
Exchanges facilitate the `movement' of trajectories from the first interface to the transition state region and back, improving the decorrelation and allowing to sample, for example, landscapes with multiple transition channels.~\cite{Bolhuis2008} 
In particular, decorrelation is improved by the stable state exchange, or minus move, where a path from the first interface is exchanged with a path exploring the stable state. Exchange moves are computationally cheap as they do not require the creation of new trajectories. The exception to this is the minus move, since new trajectories are sampled for the stable state and the first interface. Further enhancement of the exploration of the path space can be achieved by allowing exchange moves between trajectories starting in different stable states in combination with a time reversal of the trajectory. That is, a path starting in a state $A$ and ending in $B$ can be swapped with a $BA$-path. In addition, time reversal moves without swapping can further enhance decorrelation without additional computational cost.
In constrained forward shooting (CFS) RETIS,~\cite{Bolhuis2008,Bolhuis2010} shooting points are selected at the interface and, as in FFS, shooting is only performed  in the forward direction together with time-reversal moves. Path acceptance in the CFS scheme is unity and it requires smaller memory storage, but at the cost of slower decorrelation due to a reduced diversity of shooting points. The different types of moves can be combined and should be carefully tuned to achieve the highest efficiency and decorrelation for the finite number of trajectories that can be sampled for each ensemble.

The first step in every RETIS simulation is the generation of an initial path for each interface ensemble.~\cite{Bolhuis2010,Swenson2019a} There is no unique recipe to obtain a valid initial path, but generally it is a good idea to create a trial sequence of phase space points by using an MD scheme. For instance, high temperature straightforward MD can generate a trial trajectory between two states if the energy barrier can be overcome within the MD-accessible timescale. Another approach is to select a reasonable CV and employ enhanced sampling techniques, like metadynamics, to obtain a trial crossing trajectory. Quenching MD simulations or steered MD can be used as well to obtain non-equilibrium trial trajectories. 
These trial trajectories provide a collection of configurations that can serve as shooting points in an initial RETIS run to generate initial trajectories to be used for further sampling. Good initial paths quickly equilibrate and decorrelate by performing shooting/exchange/reversal moves. Poor initial paths, however, that do not follow closely enough the transition channel or the mechanism, may result in very slow decorrelation or a failure to relax toward trajectories with high probabilities in the path ensemble.
A careful check of how  the choice of initial paths affects the ensemble properties is therefore essential. In general, the final ensemble properties should be independent of the starting trajectory.

%------------------------- FFS --------------------------------------

\subsection{Forward flux sampling}

FFS is built upon the same theoretical framework as RETIS, but focuses on propagating paths only in the forward direction. Similar to RETIS, the phase space between states $A$ and $B$ is divided into smaller regions using a collection of non-intersecting interfaces along the CV $\lambda(x)$. The interface ensembles are generated sequentially using the partial paths from previous interfaces, effectively ratcheting the system towards the product state. In short, the algorithm of FFS includes ({\it i}) calculating the flux through and obtaining  configurations at the first interface $\lambda_0$ to begin the path generation, ({\it ii}) shooting several short trajectories from the current interface $\lambda_i$, and ({\it iii}) collecting the configurations that reach the next interface $\lambda_{i+1}$. Steps ({\it ii}) and ({\it iii}) of this process are repeated until the final state $B$ is reached, which is usually detected when  $P_{A}(\lambda_{i+1}|\lambda_{i})\rightarrow 1$. Thus, the two key aspects of FFS are the basin simulations, from which the first interface configurations are obtained, and the placement of the interfaces. 

In FFS, {\it all}  paths are initiated from the configurations collected at the first interface. Therefore, it is essential to ensure that the sampling at the first interface is adequate. The configurations at the first interface are a subset of the first crossings obtained from MD simulations in the basin of stable state $A$. Different approaches have been used to evaluate the convergence of the stable state sampling. First, it is important to ensure that the stored configurations at the first interface broadly sample the configurational space orthogonal to $\lambda$. Additionally, the configurations should ideally be highly decorrelated since all sampled paths will originate from them. Decorrelation is promoted by enforcing a `gap' of time between first crossings that are stored as configurations of the first interface. The `gap' is a minimal lag time calculated from the autocorrelation function (ACF) of various CVs versus number of interface crossings.~\cite{Escobedo:09:JCP} Note that for the flux calculation, {\it all} first crossings of $\lambda_0$ are considered, not just the stored crossings.

The second critical parameter in FFS is related to interface placement, since it strongly affects the sampling efficiency. On one hand, placing the interfaces too far apart results in few successful paths and  high uncertainty in the ensemble averages. On the other hand, closely placed interfaces hinder path decorrelation between interfaces and increase computational cost. Borrero and Escobedo~\cite{Escobedo:08:JCP} performed an uncertainty--cost tradeoff analysis to find the optimal interface placement or number of shoots per interface under a constrained cost. For the optimal interface placement with fixed cost (approximated as fixed $M_i$, the number of shoots from interface $i$), the interfaces should be placed such that the number of successful paths from $i$, $N_s^{(i)}$, is constant across all interfaces; that is, there should be a constant flux of trajectories at each interface. With $p_i = N_s^{(i)}/M_i$, the value of $p_i$ should be kept constant across all interfaces for the FFS calculation. $p_i$ is a free parameter that can be chosen by the user. The efficiency is not greatly impacted by the choice of $p_i$.~\cite{Escobedo:08:JCP} Two assumptions were made in this analysis. First, the cost of a trajectory is linearly proportional to its distance travelled in CV space. Second, the trajectories at different interfaces are uncorrelated. The second assumption becomes better realized as the interfaces are spaced further apart.

In general, manually placing the interfaces a priori is ill-advised. This could lead to unnecessary increases in computational cost and/or a decrease in sampling quality if too few configurations are gathered at an interface. Optimization typically leads to only a few interfaces being placed beyond the transition state region since the slope of the free energy surface tends to funnel trajectories toward the product state.

Various methods have been used for on-the-fly placement of interfaces based on the analysis of Borrero and Escobedo~\cite{Escobedo:08:JCP} outlined above. Kratzer et al.\cite{Allen:13:JCP} introduced two methods, the trial interface and exploring scouts methods, to automatically place interfaces such that the constant flux criterion is satisfied. To place $\lambda_{i+1}$ with the trial interface method, first a trial interface is proposed. A small number of trajectories is launched from $\lambda_i$ and the success probability, $p_{\text{trial}}$, is calculated. If this probability lies within acceptable bounds of the target crossing probability, $p_{\text{des}}$,  the trial interface is accepted as $\lambda_{i+1}$. Otherwise, the position of the trial interface is updated based on the difference between $p_{\text{des}}$ and $p_{\text{trial}}$. The process is repeated until the next interface is determined. 
In the exploring scouts method, $M_{\text{trial}}$ trajectories are launched from $\lambda_i$ and propagated until they reach either of the stable states or exceed $m_{\text{max}}$ steps. The trial trajectories are ranked with index $n$ based on the maximum value of $\lambda$ reached, such that $\lambda_{\text{max}}^{(n)} < \lambda_{\text{max}}^{(n+1)}$. The number of {\it unsuccessful} trajectories for a given value of $p_{\text{des}}$ is $n_{\text{des}} = [M_{\text{trial}}(1-p_{\text{des}})]$ and the interface is correspondingly placed at $\lambda_{i+1}=\lambda_{\text{max}}^{(n_{\text{des}})}$.
The selection of $m_{\text{max}}$ should balance the computational cost and exploration of larger $\lambda$ values. FFPilot~\cite{Roberts:20:JCP} is another method that optimizes the number of trajectories for each interface to reach a desired error in the rate constant. A small-scale pilot FFS calculation is performed to estimate these values prior to the full FFS calculation.

Additionally, for a jumpy CV that is capable of moving from $\lambda < \lambda_{i}$ to $\lambda \geq \lambda_{i+1}$ in a single time slice, the framework of jumpy FFS (jFFS)~\cite{HajiAkbari:18:JCP} should be used. With jFFS, paths are allowed to skip (jump) beyond two or more interfaces in a single time slice, but each encountered jump history must be accounted for in the rate calculation. As an alternative to accounting for the jump history of paths, each interface $\lambda_i$ can be placed such that it has no first crossings from $\lambda < \lambda_{i-1}$ in a single time slice.

%------------------------------ Practical guide -----------------------

\section{Practical guide}
\label{sec:practicalguide}

Motivated by our own experiences with the challenges that accompany the setup and convergence of RETIS and FFS simulations for new or increasingly complex systems, in particular for inexperienced users,  we provide a step-by-step practical guide to the key points when assessing the quality and efficiency of such simulations. This includes detecting signs of sampling issues and finding the causes of these issues, as well as identifying the parameters to adjust for achieving adequate sampling, which, in many cases, is not entirely obvious. Our goal  is to provide guidance to new users with a hands-on approach that assists them in the application of RETIS and FFS to their systems.

%----------------------------- LJ system setup ------------------------
\subsection{Example systems: Nucleation}
\subsubsection{Lennard-Jones system}

To illustrate the sampling in RETIS and FFS for a system that is more complex than simple, low-dimensional model potentials, but can still be sampled quite extensively,   
we study  crystal nucleation of Lennard-Jones (LJ) particles from the supercooled liquid. Dimensionless units are used to describe the system. $\sigma$ is the distance where the pairwise potential is zero, $\epsilon$ is the well depth, and $\tau$ is the dimensionless time unit. The particles interact with a force-switched potential from 3.0$\sigma$ to 3.5$\sigma$, as implemented in the MD software package GROMACS.~\cite{Gromacs2015} Long-range corrections are applied to the energy and pressure. A timestep of 0.001$\tau$ is used. Temperature and pressure are set to 0.8348$\epsilon/k_\text{B}$ and 5.0$\epsilon/\sigma^3$, respectively, corresponding to 78\% of the melting temperature, $T_{\text{m}}$. 
The Canonical Sampling through Velocity Rescaling (CSVR) thermostat~\cite{CSVR} and Parrinello--Rahman barostat~\cite{Parrinello-Rahman} were used to control temperature and pressure, unless otherwise indicated.
Solid particles are identified using the $q_6$ dot product of ten Wolde et al.,~\cite{Frenkel:96:JCP} calculated  between neighboring particles $i$ and $j$ as:
\begin{equation}
\label{eq:dijq6}
d_{ij}=\frac{\sum_{m=-6}^{m=6} q_{6m}(i)q_{6m}^{*}(j)}{\left(\sum_{m=-6}^{m=6} |q_{6m}(i)|^2\right)^{1/2}\left(\sum_{m=-6}^{m=6} |q_{6m}(j)|^2\right)^{1/2}}  \quad ,
\end{equation}
where
\begin{equation}
q_{6m}(i)=\frac{1}{N_n(i)}\sum_{j=1}^{N_n(i)} Y_{6m}(\textbf{r}_{ij}) \quad .
\label{eq:q6m}
\end{equation}
$N_n(i)$ is the number of neighbors of particle $i$,  $Y_{6m}$ are the $l=6$ spherical harmonics, and $\textbf{r}_{ij}$ is the  vector between particles $i$ and $j$. For a pair of particles $i$ and $j$, a solid connection is formed if $d_{ij}>0.5$. If a particle contains more than eight solid bonds, it is classified as solid. Solid nearest neighbors are then clustered together to calculate the largest cluster size, $n_s$, which is used as CV $\lambda(x)$. Particles within 1.5$\sigma$ are considered neighbors for both clustering and the  calculation of $q_{6m}(i)$. For RETIS simulations, a system size of 8192 particles was used. The systems used in FFS comprised of 16384 particles.

\subsubsection{Ni$_3$Al system}

In addition to the LJ system, RETIS was also performed to sample nucleation during solidification in Ni$_3$Al. Our analysis uses the data from Ref.~\onlinecite{Liang2020}. The system contains 6912 atoms (75\% Ni and 25\% Al). Simulations were performed at $p = 0$~bar and $T = 1342$~K (20\% undercooling) using the embedded atom method~\cite{Mishin:09:PhilMag} to model interactions and LAMMPS~\cite{lammps} as MD driver.  Further simulation details can be found in Ref.~\onlinecite{Liang2020}

%----------------------- XL hydrate system ---------------------------

\subsubsection{XL--water system}
We also performed FFS on a guest--water solution to sample the crystal nucleation of clathrate hydrates. The data from Ref.~\onlinecite{DeFever:17:JCP} is analyzed. The solution consists of water modeled by mW~\cite{Molinero:mW} and a miscible guest modeled by XL~\cite{Molinero:10:JPCB1}. Simulations were performed at a temperature of 230 K and a pressure of 500 atm using LAMMPS.~\cite{lammps} The system consisted of a cubic box with 7555 mW particles and 445 XL particles (5.6 mol\% XL).

%-------------------------- RETIS Simulations ------------------------

\subsection{RETIS simulations}

RETIS simulations require an ergodic sampling of trajectory space and its efficiency strongly depends on the number of MC moves needed to obtain converged path ensembles for each interface. A key issue is the decorrelation of successively created paths in each ensemble. Fast decorrelation will significantly reduce the computational cost, whereas slow decorrelation might indicate sampling problems, or the system might even get stuck in certain parts of trajectory space. Our guide focuses on the detection of sampling issues and checks for slow decorrelation and convergence of the path ensembles. We illustrate the change in sampling efficiency for different simulation setups studying nucleation in the LJ system.
In the cases where sampling issues are found, we proceed step-by-step to find the source of each issue and suggest changes to the parameters or algorithms to improve sampling.

\subsubsection{Computational details}
\label{sec:compdetails}

To illustrate the parameter tuning process and a range of sampling behaviors that can be encountered during RETIS, several simulations were performed with different shooting algorithms and parameters. Four different shooting algorithms were used. The first uses  two-way shooting moves with momentum perturbations, denoted NVT two-way shooting. As a the second algorithm we employ CFS,~\cite{Bolhuis2008} a one-way shooting method, included for its similarity to FFS. The third one comprises modified two-way shooting moves where the momenta are redrawn from the Maxwell-Boltzmann distribution and rescaled such that the energy remains constant. This variant is denoted here as rescaled Maxwell-Boltzmann velocities (rMBV) shooting. The fourth algorithm is similar to NVT two-way shooting, but particle momenta are rescaled after perturbation to preserve the total energy. This algorithm is denoted NVE two-way shooting.

The NVT two-way, NVE two-way, and rMBV shooting RETIS simulations use shooting and replica exchange moves. At the start of each MC move, one of the move types (shooting, exchange) is selected and is attempted for all interfaces. When, e.g., a shooting move is selected, the shooting move is attempted for the current path in each interface ensemble. The shooting point can be selected in several ways. The simplest way is to randomly select a slice from the current path with uniform probability. An alternative is to select the shooting point among all slices that lie between $\lambda_{i-1}$ and $\lambda_{i+1}$. To obey detailed balance regardless of the shooting point selection method, the maximum number of possible shooting points on the new path is set by $N^{(\text{n})}_{\text{sp,max}} = N^{(\text{o})}_{\text{sp}}/R$, where $N^{(\text{o})}_{\text{sp}}$ is the number of possible shooting points on the old path and $R \in (0,1]$ is a uniform random number.\cite{Bolhuis:03:JCP}

For NVT and NVE two-way shooting, the $x$, $y$, and $z$ components of the momenta of all particles are perturbed by a random value drawn from a Gaussian distribution with zero mean and standard deviation $\sigma_{\Delta v}$. For NVT two-way shooting, the new shooting point is accepted with the probability, $p_{acc}$:
\begin{equation}
    p_{acc} = \min\left[1,\frac{\rho(\mathbf{x}_{\text{sp}}^{(\text{n})})}{\rho(\mathbf{x}_{\text{sp}}^{(\text{o})})}\right] 
    = \min \left[1, \exp\left(-\frac{\Delta E}{k_{\text{B}}T} \right) \right]
    \quad ,
    \label{eq:momentum-acc}
\end{equation}
where $\rho(\mathbf{x}_{\text{sp}}^{(\text{n})})$ and $\rho(\mathbf{x}_{\text{sp}}^{(\text{o})})$ are the equilibrium probability densities of the new (perturbed) and old (unperturbed) shooting point configurations, respectively. In the canonical ensemble, the ratio is given by the Boltzmann factor, where $\Delta E$ is the difference in total energy between the old and new configuration. The shooting point {\it configuration} is always accepted when using NVE two-way shooting since $\Delta E = 0$.

CFS was originally designed to have an acceptance probability of unity.~\cite{Bolhuis2008} This is guaranteed by always selecting the shooting point to be the first crossing of $\lambda_i$ and only shooting in the forward direction. Like FFS, CFS requires stochastic dynamics. Here, the momenta of all particles are redrawn from the Maxwell-Boltzmann distribution at the beginning of each shooting move. The forward path is then propagated until reaching either state $A$ or $B$. To correctly capture the relaxation of the backward part of the trajectories (the path segment from state $A$ to the shooting point), CFS requires time-reversal moves in addition to the shooting and replica exchange moves. In principle, the momenta do not need to be redrawn for CFS since the stochastic dynamics should naturally lead to divergent trajectories even when starting from the same phase point. However, we found that the decorrelation of paths was not sufficiently fast for our LJ nucleation example. For FFS, the effect of momentum randomization on the crossing probabilities was evaluated for silicon nucleation. The effect was found to be negligible.\cite{Galli:09:JCP}

The rMBV shooting move is motivated by the aimless shooting algorithm~\cite{Peters06}. Instead of perturbing the momenta from a Gaussian as is done with NVT two-way and NVE two-way shooting, the momenta are redrawn from the Maxwell-Boltzmann distribution and rescaled to the kinetic energy of the unperturbed shooting point. For NVT two-way, NVE two-way, and rMBV shooting, a forward and backward path is propagated from the shooting point. The backward path must reach the initial state $A$, and the forward path must reach either state $A$ or the final  state $B$. Additionally, for interface ensemble $i$, the path must cross $\lambda_i$ and the number of possible shooting points cannot exceed $N^{(\text{n})}_{\text{sp,max}}$ for the new path to be accepted.

Our replica exchange algorithm follows that of van Erp~\cite{VanErp2007}. When a replica exchange move is selected for all interfaces, either the set of exchanges $[0^-] \leftrightarrow [0^+], [1] \leftrightarrow [2], [3] \leftrightarrow [4], ...$ or $[0^+] \leftrightarrow [1], [2] \leftrightarrow [3], [4] \leftrightarrow [5], ...$ is selected and attempted, where $[0^-]$ denotes the stable state ensemble. Here, exchanges are only attempted between neighboring interfaces.

For all RETIS simulations presented here, the overall algorithm used for every move after the initial paths have been collected is as follows. A shooting or exchange move is randomly selected for all interfaces. If a shooting move is selected, then at each interface a shooting point location on the current path is chosen, a perturbation is applied, and time integration from the shooting point is performed to generate a trial path. If a replica exchange move is selected, then the algorithm from van Erp~\cite{VanErp2007} described above is used to attempt exchanges between all adjacent pairs of interfaces.

For all RETIS simulations, slices were recorded every 0.02$\tau$. Unless noted otherwise, initial paths were generated by first simulating nucleation via straightforward MD at $0.7\,T_{\text{m}}$, where nucleation can be observed on reasonable timescales. From this simulation, configurations with $\lambda$ values at or slightly above each $\lambda_i$ were collected. The particle momenta of each configuration were reassigned from the Maxwell-Boltzmann distribution at $0.78\,T_{\text{m}}$ and propagated forward and backward in time. This was repeated until a valid initial path, starting in state $A$ and ending in either $A$ or $B$, was obtained for each interface.

%---------------------------- Initial diagnostic ------------------------------------

\subsubsection{Initial diagnostic plots and expected behavior}

The performance and accuracy of the rate constant calculation with RETIS is determined by two key factors: ({\it i}) the quality of the CV chosen to discriminate between the stable states and ({\it ii}) the quality of the sampling of trajectories within the MC approach. 
Adequate and efficient sampling of the path ensemble requires fast decorrelation of the paths and a high acceptance ratio within the MC framework, which guarantees and enhances ergodicity. In addition, the computation of the rate constant $k_{AB}$ and statistically relevant transition mechanisms is intrinsically associated with an accurate estimation of the reactive flux from $A$ to $B$, defined by the interface crossing probabilities. Thus, a comprehensive inspection of the interface crossing probabilities provides valuable information about the quality of the path ensemble. Such close analysis can very often guide the detection and resolution of sampling problems, as well as the fine-tuning of parameters that lead to an adequate and efficient sampling of the path ensemble. 

\begin{figure*}
\centering
\includegraphics[width=0.9\linewidth]{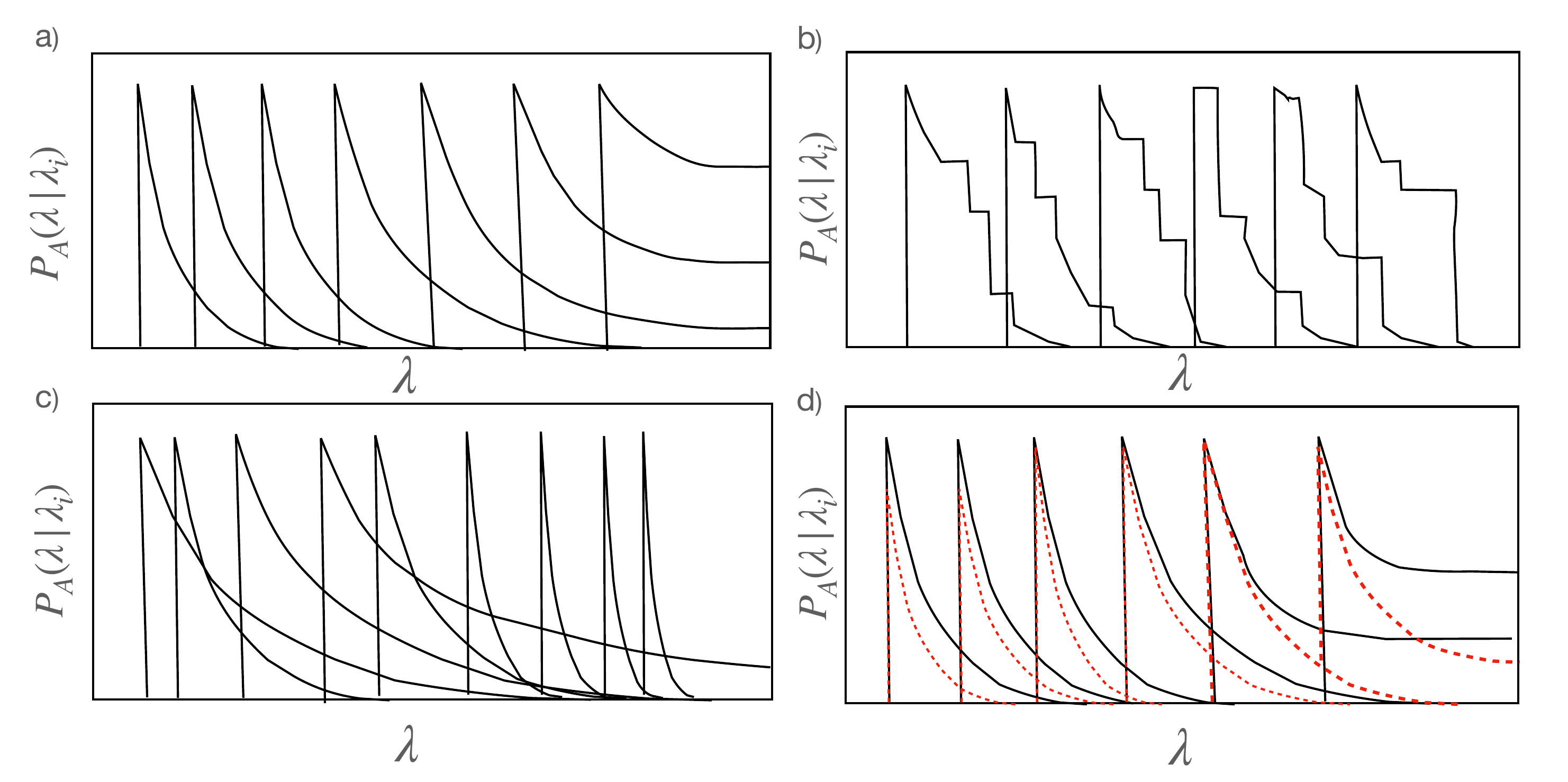}
\caption{Illustration of the expected behavior and possible pathological behaviors of the crossing histograms. a) Well-behaved crossing histograms as a function of  $\lambda$. b) Typical case of ill-behaved and noisy crossing histograms that exhibit unexpected plateaus. c) Example of crossing histograms, where the probability of interface $\lambda_i$ exceeds the one of $\lambda_{i+1}$ for larger $\lambda$ values. d) A case of crossing histograms that are well behaved in appearance (black lines), but do not converge when increasing the sampling of paths or for independent ensembles with the same number of  paths (red dashed lines). }
\label{fig:schematics}
\end{figure*}

Based on our hands-on experience with RETIS in complex systems,~\cite{DiazLeines2017,DiazLeines2018,Liang2020,Menon2020} in Fig.~\ref{fig:schematics} we summarize the ideal behavior of the interface crossing probabilities and common ill-behaved crossing histograms encountered due to sampling problems. The expected behavior of the crossing histograms for all interfaces from $A$ to $B$, placed at different values of the CV, $\lambda$, is shown in Fig.~\ref{fig:schematics}(a). The crossing histograms are smooth with sufficient overlap (at least 10\%, but commonly 15-20\% is desired). The probability to reach state $B$ or larger values of $\lambda$ is naturally very small at interfaces placed close to stable state $A$. As $\lambda$ approaches and crosses the transition state region, the probability to reach state $B$ is expected to increase. The crossing probabilities tend to increase towards unity  (Fig.~\ref{fig:schematics}(a)) as the majority of  trajectories commits to state $B$. 

The first and most clear indication of sampling problems is shown in Fig.~\ref{fig:schematics}(b). The crossing histograms are noisy and exhibit unexpected plateaus. Noisy behavior can simply indicate insufficient sampling, which results in crossing histograms that are not yet converged. That is, the number of paths collected is not statistically representative of the ensemble and more MC moves are required. The presence of plateaus in the crossing histograms usually indicates that many paths are repeatedly sampled due to low acceptance. This hints at a problem of high path correlation in the ensemble, especially if the plateaus do not smooth out and disappear with more sampling.

A less obvious case of ill-behaved crossing histograms is illustrated in  Fig.~\ref{fig:schematics}(c). The histograms appear smooth and overlap, but the crossing probabilities do not display the expected behavior. The probability to reach state $B$ or larger values of $\lambda$ is low at interfaces supposedly near the transition state region or already in the basin of attraction of state $B$.  Furthermore, there is a higher probability to reach state $B$ or larger values of $\lambda$ at interfaces placed near state $A$, compared to the interfaces closer to state $B$. This problem generally indicates that the interface ensembles have not yet relaxed or decorrelated from the initial path. A poor choice for the generation of the initial trajectory, e.g., by abruptly taking the system out of equilibrium, can result in such slow decorrelation. This behavior also implies that the replica exchange moves might not be well-tuned, which should, in principle, alleviate biased crossing probabilities as observed here. In the third scenario, Fig.~\ref{fig:schematics}(d), we illustrate a problem of convergence of the crossing histograms. Although the crossing probabilities appear smooth and well-behaved (black lines in Fig.~\ref{fig:schematics}(d)), statistically independent ensembles with an equal number of paths display different crossing probabilities.  Convergence of the histograms cannot be achieved in any reasonable amount of time, even if the number of paths in the ensembles is increased. This behavior is most often observed when the transition state region and the stable states are not well-distinguished. This indicates that the CV used for $\lambda$ provides a poor approximation of the committor probability.

The inspection of the crossing histograms is a very informative and simple way to detect sampling issues, especially for complex systems and should be included in the best practices for RETIS simulations.  However, to determine the root causes of sampling problems and their resolutions, other rigorous tests need to be performed. Such tests include probing the acceptance ratios, ACF of paths, replica mobility through interface ensembles, an analysis of the diversity of the shooting points and paths, sampling trees, path length distributions, etc.~\cite{Dellago2002,Bolhuis2015}

\begin{figure}[h]
\centering
\includegraphics[width=\linewidth]{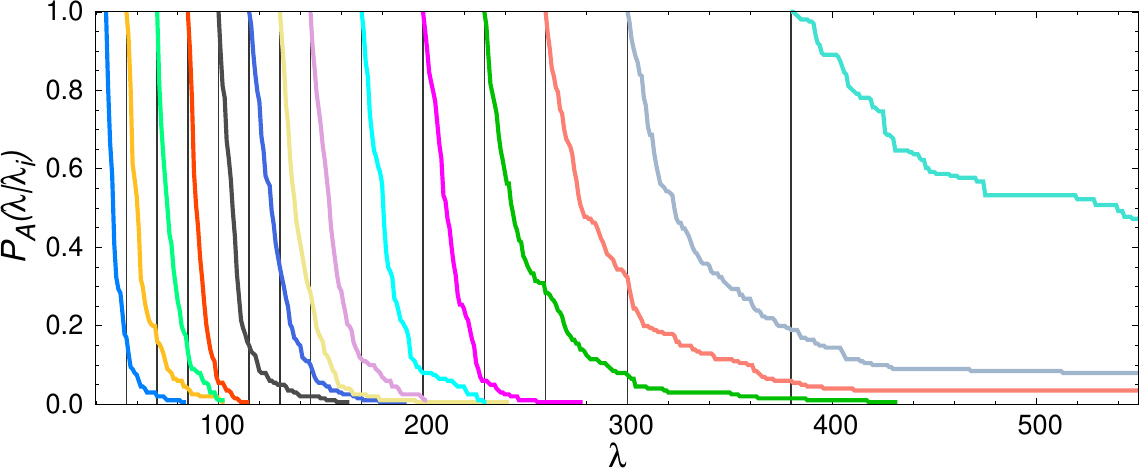}
\caption{Example of nearly ideal crossing histograms obtained from RETIS of LJ crystal nucleation. The crossing histograms were collected using 50\% rMBV shooting moves and 50\% replica exchange moves. Each shooting point was selected to lie between $\lambda_{i-1}$ and $\lambda_{i+1}$. Every fifth path from move 800--3000 was used, yielding a total of 441 paths.}
\label{fig:aimlesshist-full}
\end{figure}

\subsubsection{Rate constant calculation using RETIS}

The estimation for the rate constant of crystal nucleation in a LJ system illustrates an application of RETIS in complex rare event transitions.  In Fig.~\ref{fig:aimlesshist-full} we show the converged crossing histograms that  exhibit the expected trends, indicating an appropriate and sufficient sampling needed for an accurate calculation of the rate constant. The exception may be the final interface ensemble, which appears quite rough and is still decreasing slightly at the boundary of the solid stable state. We performed RETIS simulations using the largest solid cluster size $n_s = \lambda$ as the CV. The positions of the interfaces were carefully selected such that there is at least $10\%$ overlap between $\lambda_i$ and $\lambda_{i+1}$ (Fig.~\ref{fig:aimlesshist-full}). Ensembles of paths were collected at interfaces positioned at $\lambda_i = \{$45, 55, 70, 85, 100, 115, 130, 145, 170, 200, 230, 260, 300, 380$\}$. The liquid state region is defined as $\lambda < 45$ and the boundary of the solid stable state is set at $\lambda \geq 550$. The boundary of the solid state was chosen such that $>95$\% of  trajectories with $n_s > 550$ commit to the solid state. No sampling was performed for the solid stable state. After careful  diagnostic tests of the sampling quality and subsequent tuning of parameters to improve sampling, we performed RETIS simulations with a rMBV shooting algorithm, including 50\%  moves and 50\% replica exchange moves. We performed a total of 3000 moves, which is found sufficient for the calculation of the rate constant. The collection of paths included in the histogram calculation began after 800 moves to ensure that bias from the initial paths had been removed. Paths were collected every fifth move to enforce decorrelation between successive paths. This also decreases the storage cost without compromising the accuracy of ensemble averages over the trajectories.

\begin{figure*}
\centering
\includegraphics[width=0.8\linewidth]{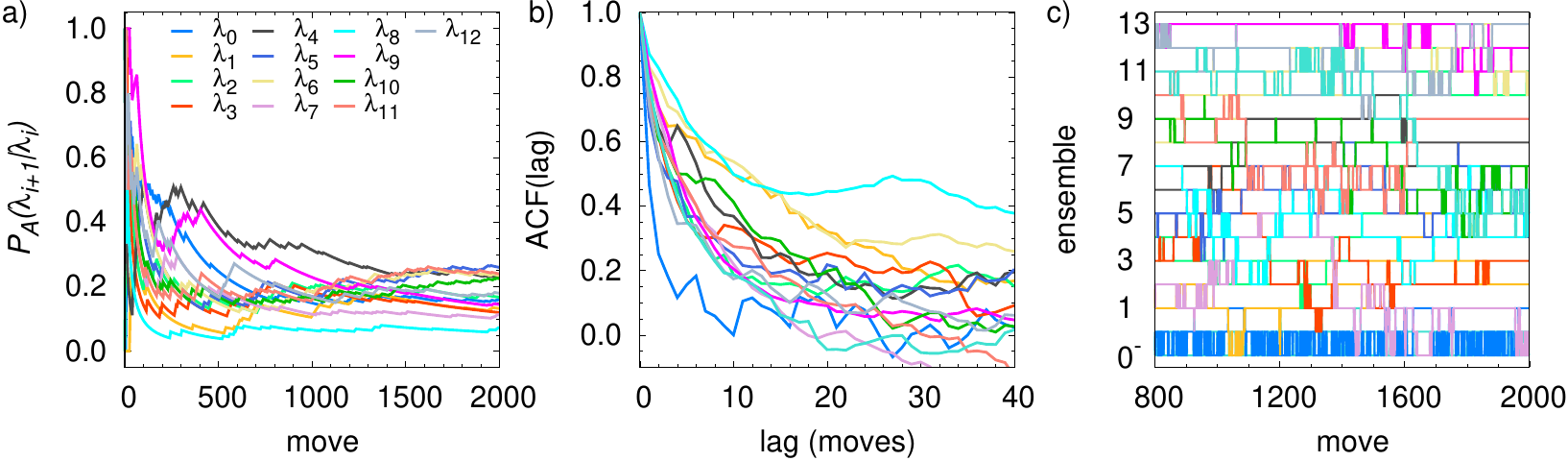}
\caption{Diagnostic plots for the rMBV shooting RETIS simulations with 50\% rMBV shooting moves and 50\% swapping moves: (a) running average of crossing probabilities, (b) path length ACF, and (c) path mobility. The ACF in (b) was calculated using paths collected from moves 800--2000.}
\label{fig:aimlessdiagnostics-full}
\end{figure*}

This yielded a total of 441 paths per interface ensemble from which the crossing histograms shown in Fig.~\ref{fig:aimlesshist-full} were calculated. The final interface crossing histogram would ideally plateau prior to reaching $\lambda_B$, as is observed with the two interfaces just before it. This lack of plateau is an indication that $\lambda_B$ should ideally be placed at a higher value to obtain a more accurate estimate of the rate constant. The trade-off is that more time will be required to sample trajectories that reach $\lambda_B$. We expect that the roughness is caused by a shooting move acceptance probability that is approximately half that of the other interfaces. This occurs due to the generation of $B$-to-$B$ paths, which are always rejected.

The sampling quality is assessed with diagnostic plots shown in Fig.~\ref{fig:aimlessdiagnostics-full}. The running average of crossing probabilities for all interfaces, shown in Fig.~\ref{fig:aimlessdiagnostics-full}(a), provides a measure of the number of moves needed to converge the crossing histograms. This is calculated by taking the average of the crossing probability from move zero to the current move. The average crossing probabilities appear to converge by move 800, implying that we need to include at least 800 moves for adequate relaxation of the path ensembles. To assess the decorrelation of the paths in RETIS, we compute 
the ACF of a scalar property of the path:~\cite{Dellago2002}
\begin{equation}
    \mathrm{ACF}(\mathrm{lag})=\sum_{i=1}^{N-\mathrm{lag}} \frac{\left(\theta_{i}-\overline{\theta}\right)\left(\theta_{i+\operatorname{lag}}-\overline{\theta}\right)}{\sum_{i=1}^{N}\left(\theta_{i}-\overline{\theta}\right)^{2}}.
\end{equation}
$N$ is the number of moves, $\theta_i$ is a property of the path at move $i$, and $\overline{\theta}$ is the average value of $\theta$ across all moves. The ACF provides an estimate of the number of moves needed  such that each path is decorrelated from the previous sample.
Commonly, the slowest-decorrelating property that is relevant to the transition should be considered in the analysis to guarantee a proper assessment of the path decorrelation.  $\theta_i$ can also be the binary indicator function of whether or not a path crosses $i+1$---so $\theta_i=0$ if a path does not cross the next interface and $\theta_i=1$ if it does. This has the advantage of being closely related to the uncertainty of the crossing probabilities.\cite{vanErp:06:JCP} In Fig.~\ref{fig:aimlessdiagnostics-full}(b) we show the ACF of the path length, which is found to provide a good estimate of the number of moves required to sample independent paths. Fig.~\ref{fig:aimlessdiagnostics-full}(b) shows that paths are decorrelated after approximately five moves for all the interfaces and thus collecting the paths after five steps was found as a good trade-off between computational time of sampling and appropriate decorrelation of the path ensemble. The number of moves between independent paths can also be estimated by calculating the autocorrelation time, which is
\begin{equation}
    \tau_{\mathrm{ACF}} = \sum_{\mathrm{lag}=1}^{\infty}{\mathrm{ACF}(\mathrm{lag})}.
    \label{eq:tau-acf}
\end{equation}
The swapping quality can, in general, be evaluated by computing the replica flow and replica round trip times.~\cite{Swenson2019b}  However, these quantities are rarely accessible in finite sampling of complex systems.
Here, the swapping quality is assessed by visualizing the replica mobility (Fig.~\ref{fig:aimlessdiagnostics-full}(c)) along interfaces.~\cite{Bolhuis2008} Ideally, effective replica exchange is indicated by replicas diffusing throughout the entire range of interfaces, exploring efficiently the transition channel. 
While most replicas appear to move through a few neighboring interfaces well, some sparsity of swapping is observed among the interfaces. This sparsity could result from the combination of low crossing probabilities and an insufficient number of moves compared to the number of interfaces present. The resulting shooting move acceptance probability in our RETIS simulation is at least 35\% for each interface, except for the final one, which is an optimal acceptance to achieve appropriate decorrelation of the path ensemble.~\cite{Dellago2002}

\begin{figure}
\centering
\includegraphics[width=0.6\linewidth]{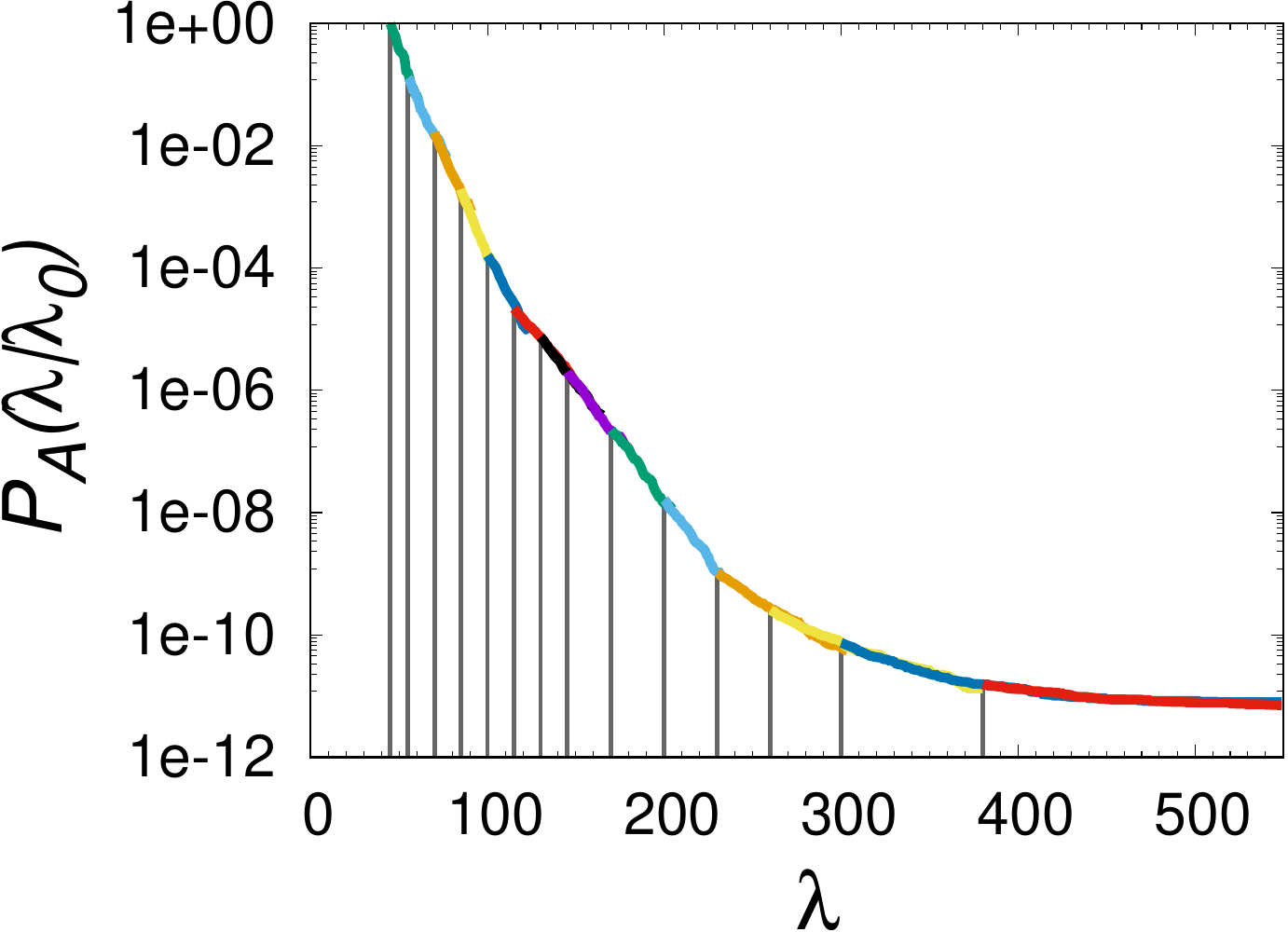}
\caption{
Matched crossing histograms obtained from the rMBV RETIS simulation.}
\label{fig:mbv-full}
\end{figure}

The rate constant $k_{AB}$ and corresponding  nucleation rate  $ J=k_{AB}/V$, where $V$ is the average volume of the simulation box containing the liquid phase, were calculated. The flux $\Phi_0$ was calculated from the average lengths of paths in the [0$^-$] and [0$^+$] ensembles. $P_{A}(\lambda_n|\lambda_0)$ was obtained from matching the crossing histograms for all interfaces employing the weighted histogram analysis method (WHAM)~\cite{Ferrenberg1989} with a cutoff probability of 0.05, which is the minimum value of $P_{A}(\lambda | \lambda_i)$ for which interface $i$ will contribute to the overall crossing histogram. The resulting matched histograms are shown in Fig.~\ref{fig:mbv-full}.
The uncertainty in $P_{A}(\lambda_n|\lambda_0)$ was estimated with bootstrapping. 500 bootstrap samples were collected, each consisting of 441 paths randomly sampled with replacement from each interface.
The resulting nucleation rate is  $\log_{10}(J_{\text{RETIS}})=-16.6 \pm 0.51$ for a 95\% confidence interval. This is in agreement with the rate obtained from our FFS simulations, $\log_{10}(J_{\text{FFS}}) = -16.2 \pm 0.44$, as discussed in Section~\ref{subsec:ffs}.

\subsubsection{Assessing the sampling quality}

In the previous section, we demonstrated that RETIS provides an accurate and efficient estimation of rate constants in complex systems if the path ensemble is sufficiently converged. To arrive at such a result, however, it is necessary to assess the sampling quality of RETIS simulations and tune parameters to improve it. To this end, we show examples of various ill-behaved path ensembles that we have encountered during our study of the LJ system and which illustrate common scenarios of sampling problems (see Fig.~\ref{fig:schematics}). We discuss practical guidelines to detect and tune  RETIS simulations in order to achieve better sampling and convergence. The effects of various parameters on the sampling quality are first presented separately. However, it is very much possible that, in practical cases, several factors simultaneously hinder an efficient sampling. This will be briefly addressed towards the end of the section.

\begin{figure}[h]
\centering
\includegraphics[width=0.6\linewidth]{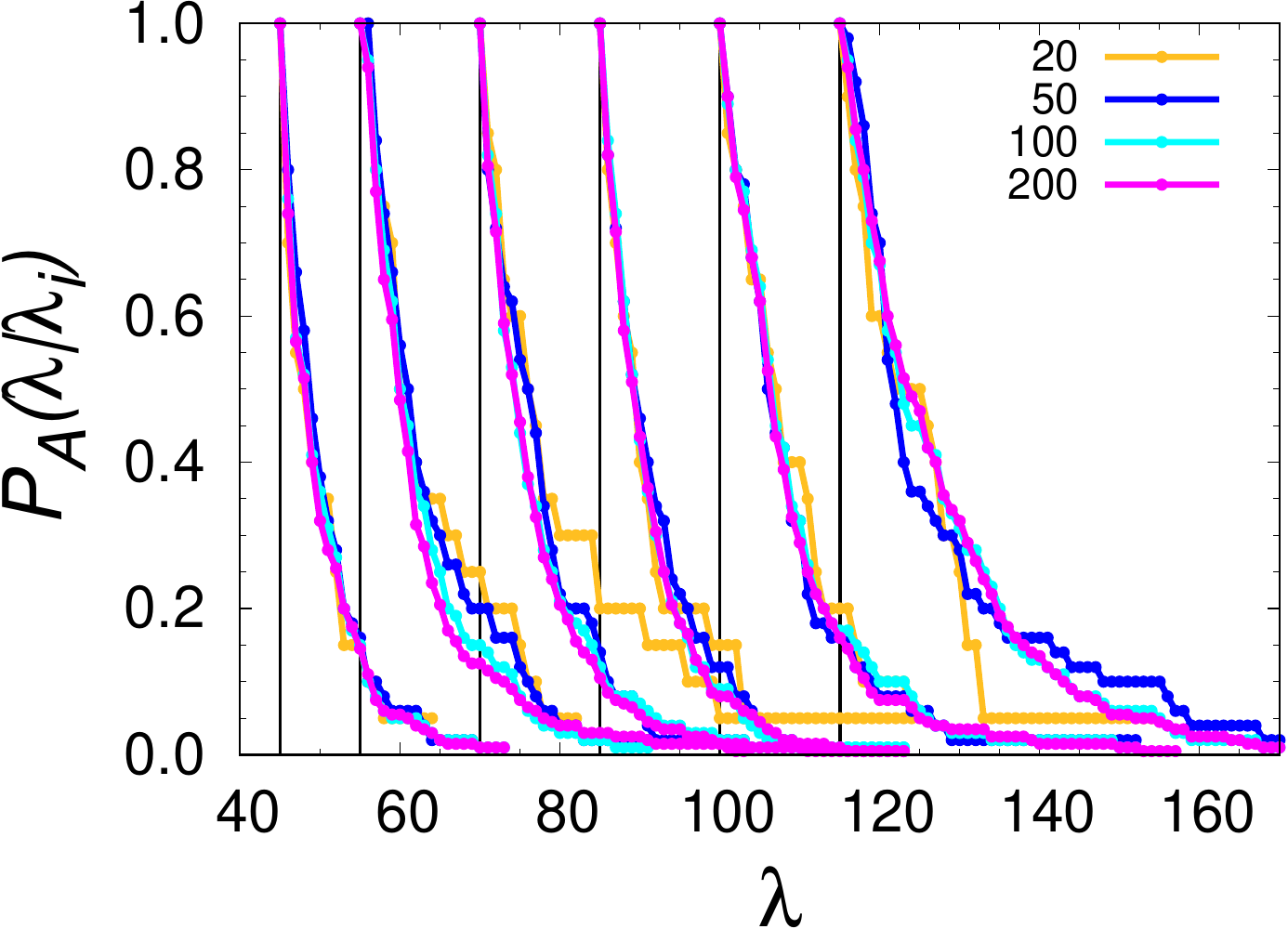}
\caption{Dependence of the rMBV shooting RETIS crossing histograms on the number of paths. Each crossing histogram was calculated using the number of paths indicated in the key. Each set of paths was evenly distributed over the range of moves 800--2000.}
\label{fig:aimlesshist-nummoves}
\end{figure}

\paragraph{Noisy crossing histograms}
The first common type of sampling problems can be identified from crossing histograms that are noisy and/or have large plateaus, like those shown in Fig.~\ref{fig:schematics}(b). This is typically related to insufficient sampling of decorrelated paths. One possible cause is that simply not enough paths were included in the statistical ensemble, yielding the noisy appearance. An example of this is illustrated in Fig.~\ref{fig:aimlesshist-nummoves} for the interface ensembles of the rMBV shooting RETIS simulation. The crossing histograms are calculated including different number of paths in the ensemble. For each set of paths, consecutive paths have equal number of moves between them, and the paths span the range of moves 800--2000. We only include paths that are decorrelated to ensure that statistical noise is only related to the ensemble size.
Fig.~\ref{fig:aimlesshist-nummoves} shows that including 20--50 paths in the ensemble results in rather noisy crossing histograms, while
including 100 decorrelated paths is sufficient to reach convergence, indicated by smooth crossing histograms that do not change significantly when increasing the ensemble size. 
Note that an ensemble size of 50 decorrelated paths appears to be adequate for a preliminary estimate of the crossing histograms and overlap. This allows for an initial inspection and subsequent tuning of the interface placements without sampling the number of paths needed for full convergence.

\begin{figure}
\centering
\includegraphics[width=0.95\linewidth]{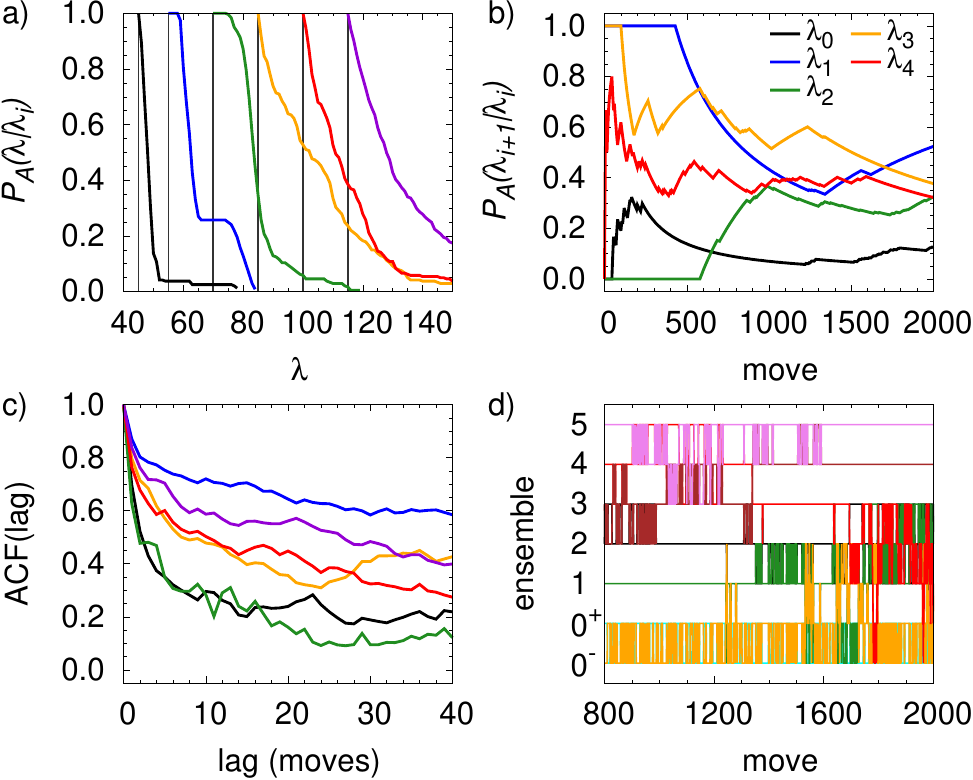}
\caption{Diagnostic plots for the NVT two-way shooting RETIS simulations. (a) Crossing histograms, (b) running average of crossing probabilities, (c) path length ACF, and (d) path mobility. The color code for panels (a), (b), and (c) is indicated in the key in panel (b). Every fifth path of moves 800--2000 were used to calculate the crossing histograms. The ACF was calculated with all paths from moves 800--2000.}
\label{fig:nvt-hist-cp}
\end{figure}

If increasing the size of the path ensemble does not resolve the noise and plateaus of the crossing histograms, then it is likely that the paths are highly correlated. In Fig.~\ref{fig:nvt-hist-cp}(a) we show an example of this case using  NVT two-way shooting and a path saving frequency of every 5 paths. The first three interfaces in particular display prominent plateaus, an indication that the same or very similar paths are being repeatedly sampled. Moreover, to assess the convergence of the crossing histograms, in Fig.~\ref{fig:nvt-hist-cp}(b) we show the running average of the crossing probabilities $P_{A}(\lambda_{i+1} | \lambda_i)$  as a function of the number of MC moves. We observe that convergence is not reached even after 2000 MC moves. Furthermore, there are long stretches where either almost every path crosses the next interface for $\lambda_1$ (blue curve, moves 1300--2000), or almost none for the same interface (moves 500--1300). These are indicated by the monotonically increasing and decreasing sections of the curve, respectively. Instead, we expect to observe a convergence behaviour similiar to the one shown in Fig.~\ref{fig:aimlessdiagnostics-full}(a). Additionally, the path length ACF (Fig.~\ref{fig:nvt-hist-cp}(c)) decays very slowly for most of the interfaces, and several do not decorrelate within 40 moves. The crossing histograms in Fig.~\ref{fig:nvt-hist-cp}(a) were calculated using every fifth move, whereas the ACF indicates we may need more than 40 moves between subsequent uncorrelated paths. In such a case, the decorrelation of the paths is cleary poor compared to the rMBV shooting case presented in Fig.~\ref{fig:aimlessdiagnostics-full}(b). 

High path correlation is very often related to the selection and generation of shooting points and  poor replica exchange. In our example, we first examine each aspect of the shooting moves to understand the causes of slow path decorrelation. Other possible causes of slow decorrelation will be discussed later. The type and magnitude of the shooting point perturbations are central components in achieving fast path decorrelation and sufficient acceptance, which are key aspects for an ergodic sampling of the pathways in TIS simulations. If the magnitude of the perturbation is small, the acceptance of new paths will be large, but subsequent paths will be highly correlated. If the perturbation magnitude is  large, accepted paths will be less correlated, but the acceptance probability will be low, leading to repeated sampling of the same paths. 

\begin{figure}[h]
\centering
\includegraphics[width=0.95\linewidth]{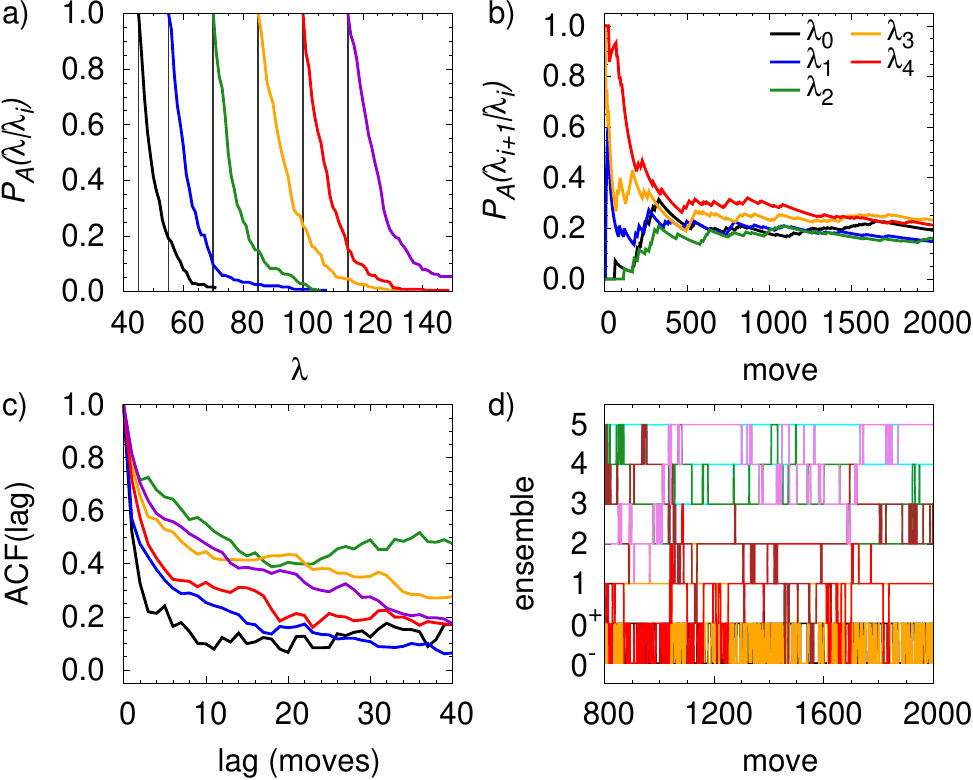}
\caption{Diagnostic plots for the NVE two-way shooting RETIS simulations: (a) crossing histograms, (b) running average of crossing probabilities, (c) path length ACF, and (d) path mobility.}
\label{fig:nvediagnostics}
\end{figure}

In general, the perturbation is tuned to yield an overall acceptance of shooting moves of 40--50\%.  Comparing NVT and NVE two-way shooting, we observe, however, significant differences.  As shown in Table~\ref{tab:momentumrejection}, the magnitude of the perturbation has to be chosen much smaller in the case of NVT two-way shooting ($\sigma_{\Delta v} = 0.05$) compared to NVE two-way shooting ($\sigma_{\Delta v} = 20$).  The reason for this is that most of the shooting moves are already rejected at selection of the shooting point configuration according to Eq.~\eqref{eq:momentum-acc}.  This becomes particularly problematic for large system sizes as the total energy difference before and after the perturbation enters Eq.~\eqref{eq:momentum-acc}.  Consequently, only very small perturbations are possible, leading to high correlation between trajectories in the ensemble.
For the LJ system, NVE two-way shooting allows for much larger perturbations and significantly improves the decorrelation and convergence of the path ensemble, as shown in Fig.~\ref{fig:nvediagnostics}.

\begin{table}
\linespread{0.9}\selectfont\centering
\caption{Fraction of rejected shooting moves that were rejected at the momentum perturbation step during the NVT two-way shooting RETIS simulation ($f_{\text{rej,NVT}}^{\text{momentum}}$) and overall shooting move acceptance probabilities for the RETIS runs using NVT and NVE two-way shooting moves ($p_{\text{acc,NVT}}$ and $p_{\text{acc,NVE}}$). $f_{\text{rej,NVT}}^{\text{momentum}}$ is calculated as the number of shooting moves that were rejected at the momentum perturbation step divided by the total number of rejected shooting moves. For NVT two-way shooting, $\sigma_{\Delta v}=0.05$. For NVE two-way shooting, $\sigma_{\Delta v}=20$. $\sigma_{\Delta v}$ is the standard deviation used to draw momenta perturbations from a zero-mean Gaussian distribution.}
\label{tab:momentumrejection}
\begin{center}
\setlength{\tabcolsep}{6pt}

\begin{tabular}{@{} l c c c @{}} \toprule
$i$ & $f_{\text{rej,NVT}}^{\text{momentum}}$ & $p_{\text{acc,NVT}}$  & $p_{\text{acc,NVE}}$\\
\midrule
0  &  0.449 &  0.455  &  0.326 \\
1  &  0.476 &  0.483  &  0.374 \\
2  &  0.840 &  0.466  &  0.630 \\
3  &  0.736 &  0.517  &  0.517 \\
4  &  0.746 &  0.566  &  0.542 \\
5  &  0.760 &  0.536  &  0.561 \\
\bottomrule

\end{tabular} 
\end{center}  
\end{table}

Another aspect that significantly impacts the path correlation is the diversity of configurations selected as shooting points. To achieve proper path decorrelation and ergodicity, not only is the magnitude of the shooting point perturbation relevant, but shooting moves must also ensure the sampling of a variety of shooting point configurations that yield diverse pathways while guaranteeing appropriate acceptance probability. Both aspects, the magnitude of the perturbation and the diversity of shooting configurations, influence the acceptance probability and path decorrelation, and thus the quality of the sampling.
One of the most straightforward shooting point selection methods is to randomly pick a slice from the entire path. While the resulting diversity of shooting configurations is large, especially for long paths, this selection leads to an exceptionally low acceptance probability.
Instead, the shooting points can be biased to lie near $\lambda_i$ (Appendix 3 of Ref. \onlinecite{Bolhuis2008}). One way of accomplishing this is to restrict the shooting point selection between $\lambda_{i-1}$ and $\lambda_{i+1}$, or more strictly, selecting only the first crossing of $\lambda_i$. The latter, used for example in CFS, guarantees high acceptance of the trial path. This allows greater perturbations to the shooting point, which aid in path decorrelation. Selection of the first crossing potentially comes at the cost of lowering the diversity of shooting point configurations. Though a high acceptance of trial paths is expected, the change between subsequent shooting points, and therefore paths, may be small, especially if the path is generated only in the forward direction, as is done with CFS.

In the following we compare two scenarios where the diversity of shooting points impacts the quality of the sampling of the interface ensembles. To this end, we used two shooting point selection methods, namely rMBV and CFS, and performed RETIS simulations. For simulations with rMBV shooting moves the initial paths were collected from move 3000 of the rMBV shooting RETIS simulation shown in Fig.~\ref{fig:aimlesshist-full}. Only the first six interfaces were used,  a total of 1000 moves were performed, and paths were saved every fifth move. For CFS RETIS the shooting point was selected as the first crossing of the interface and the trial trajectory was only propagated in the forward direction from the shooting point, leaving the backward part of the path unmodified. Here the initial paths were obtained from a nucleation trajectory simulated at a lower temperature, as described at the end of Sec.~\ref{sec:compdetails}. 

\begin{figure}
\centering
\includegraphics[width=\linewidth]{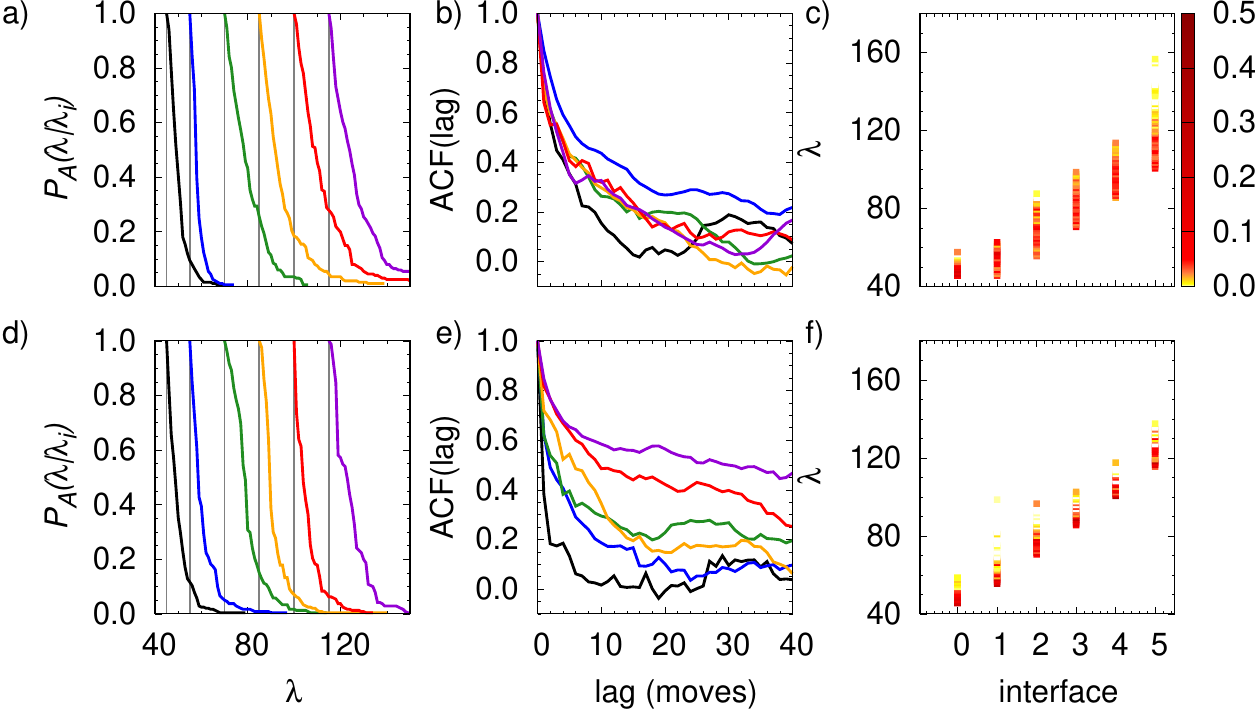}
\caption{Crossing histograms, path length ACFs, and shooting point distributions for rMBV RETIS with shooting point selection between $\lambda_{i-1}$ and $\lambda_{i+1}$ (a--c), and with CFS RETIS (d--f).}
\label{fig:sp-hist-acf}
\end{figure}

The diagnostic plots for two shooting moves are shown in Fig.~\ref{fig:sp-hist-acf}. Selection of the first crossing as the shooting point with CFS shows crossing histograms (Fig.~\ref{fig:sp-hist-acf}(d)) that look fairly smooth with small kinks and non-idealities---e.g., at interfaces 2, 4, and 5. Nevertheless, slow decorrelation of the ACFs (Fig.~\ref{fig:sp-hist-acf}(e)), especially for the higher $\lambda$ interface values, indicates poor sampling quality. In contrast, with the shooting point selection between $\lambda_{i-1}$ and $\lambda_{i+1}$ in rMBV RETIS, the crossing histograms are fairly reasonable and faster decorrelation of the ACFs is achieved (Fig.~\ref{fig:sp-hist-acf}(a and b)). The source of slow decorrelation of the ACFs may be identified in the diversity of shooting points. The diversity of shooting points can be assessed by calculating the distribution of shooting points in the $\lambda$-space. The ideal scenario is to have the broadest range of shooting points possible at each interface to enhance ergordic sampling, while maintaining an adequate shooting move acceptance probability. From Fig.~\ref{fig:sp-hist-acf}(c and f), we see that the shooting point selection algorithm of rMBV RETIS leads to a broader distribution of $\lambda$ values at every interface, which we expect to lead to a greater diversity of shooting configurations and paths. Across all interfaces, the shooting move acceptance probability of rMBV shooting lied within a narrow range from 0.46 to 0.57, which is acceptable. A further calculation revealed that only 25\% of the shooting points from CFS RETIS were unique, compared to 68\% from rMBV RETIS. This indicates that the slower decorrelation is likely a result of a low diversity of shooting points.

Selecting the first crossing point as shooting point is however not always a cause of low diversity of shooting points and poor sampling, and it can be beneficial for improving the acceptance.~\cite{Bolhuis2008} A combination of first crossing point selections with other shooting moves like rMBV RETIS may provide enough new crossing points and allow for proper sampling.
Indeed, rMBV RETIS with a first crossing point selection of shooting points showed much faster decorrelation of the paths (not shown) than CFS RETIS and did not display noticeable sampling issues. As a rule of thumb for achieving adequate sampling, it is always desirable that the selection of shooting points and shooting moves provide the highest diversity of shooting points while allowing for proper acceptance.

So far, the selection of shooting points and the size of the momentum perturbation have only been discussed in isolation, but these two aspects are interdependent. For example, the selection of first-crossing shooting points allows larger momentum perturbations, but can lead to low diversity of shooting configurations and thus a slower path decorrelation. However, the use of larger perturbations can in other cases counteract this effect and allow for adequate diversity of shooting points and paths. The interdependence of RETIS parameters and their effects on the quality of the sampling ultimately depend on the system and the dynamics under study. Nevertheless, these diagnostic tests provide general guidelines for finding the common sources of high path correlation and bumpy crossing histograms, which can direct the subsequent fine tuning of these contributions.

Another indication of sampling problems is a poor mobility of replicas between the interfaces, as shown in Fig.~\ref{fig:nvt-hist-cp}(d) for the NVT two-way shooting RETIS simulation.
There are several interfaces that experience long periods of no swapping in either direction (e.g., $i=1$ for moves 800--1200), or where swapping only occurs in one direction. This behavior indicates slow path decorrelation and it is important to note that poor replica mobility is most likely a symptom rather than a cause. Therefore, the shooting move and interfaces should be reasonably tuned before tuning the replica exchange parameters.

\begin{figure}
\centering
\includegraphics[width=\linewidth]{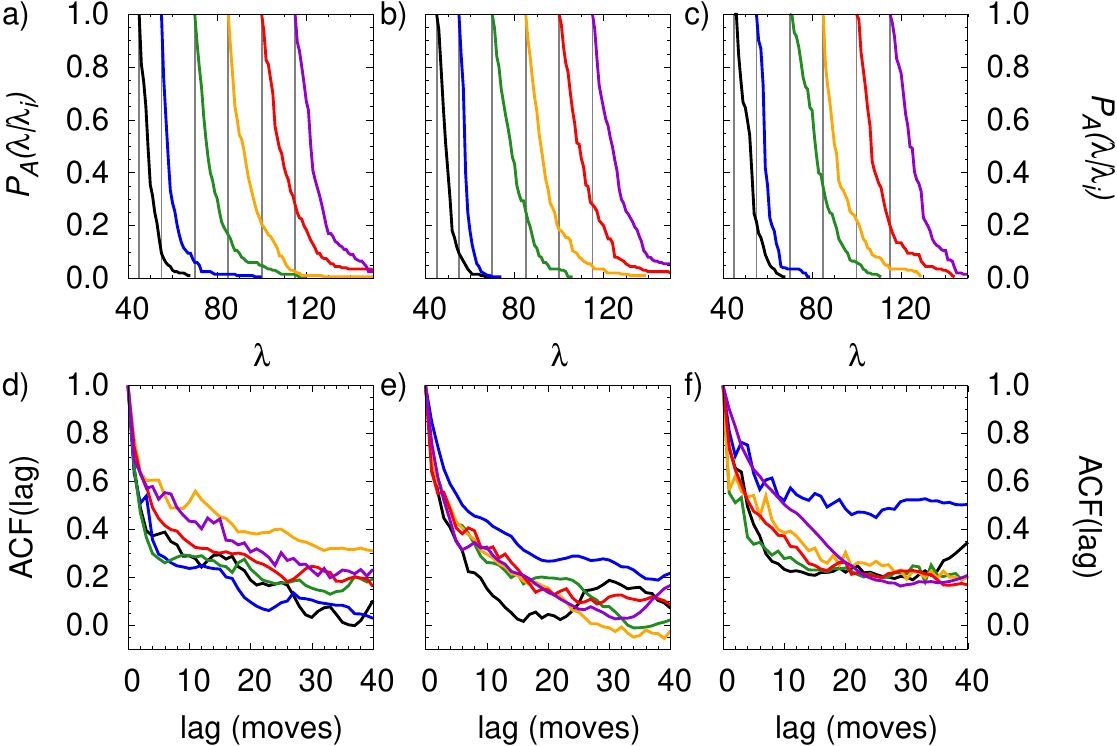}
\caption{Effect of the ratio of swapping moves on the crossing histograms (a--c) and path length ACF (d--e). The percentage of swapping moves is 0\% (a and d), 50\% (b and e), and 80\% (c and f).}
\label{fig:swapratio}
\end{figure}

Once the shooting moves are well-tuned, swapping moves aid in decorrelating paths efficiently. The optimal ratio of shooting and swapping moves can be tuned by evaluating the crossing histograms and ACFs. To explore this effect, rMBV shooting move RETIS simulations were performed with no swaps, 50\% swaps, and 80\% swaps. 1000 moves  were performed in total, and every fifth path was stored yielding a total of 200 paths included in the analysis. Fig.~\ref{fig:swapratio} shows the resulting crossing histograms and path length ACFs for these simulations. In this example, a comparison of the crossing histograms in Figs.~\ref{fig:swapratio}(a and b) indicates that simulations with no swapping and 50\% swapping moves yield similar results. This is corroborated by the  ACFs in Figs.~\ref{fig:swapratio}(d and e). With 80\% swaping moves (Fig.~\ref{fig:swapratio}(c)), the crossing histograms exhibit a rougher appearance and decorrelate slower than in the other two cases (Fig.~\ref{fig:swapratio}(f)), indicating an overall lower quality of the sampling. The  cause of this behavior is that swapping moves do not generate new trajectories and thus, the same paths are being swapped between interfaces many times. This emphasizes the importance of having a sufficient number of accepted shooting moves to guarantee the diversity and decorrelation of the paths. 

The main advantage of swapping moves is that they are computationally cheap as they do not require the creation of new trajectories (except for swapping moves with the stable state ensemble).  Consequently, swapping moves can rather easily be added to enhance decorrelation while keeping the total number of shooting moves constant.

In this section, guidelines to identify potential causes of noisy crossing histogram and slow path decorrelation have been introduced.  In the following section, convergence problems due to the choice of $\lambda$ are discussed.

\paragraph{Non-convergence and improper convergence of crossing histograms.} 
Even if a large number of decorrelated paths is obtained, it might still not be possible to converge the crossing histograms within any reasonable amount of sampling. Crossing histograms can appear smooth at first sight but exhibit ill-behaviour---e.g., suspiciously high crossing probabilities for interfaces close to the stable states (see Fig.~\ref{fig:schematics}(c)). Such behavior of the crossing histograms often hints towards additional sampling problems associated with ``trapping'' of pathways in regions away from the transition region, yielding a poor approximation of the committor probability, which is ultimately related to the crossing probability.
This can potentially be caused by poor selection of the initial paths for the interfaces, leading to paths which are far off from the transition tube that we wish to sample. Another example is the selection of a poor $\lambda$ for the sampling CV that cannot capture the transition region appropriately. With finite sampling, both issues result in an incorrectly sampled TPE.

\begin{figure}
\centering
\includegraphics[width=0.9\linewidth]{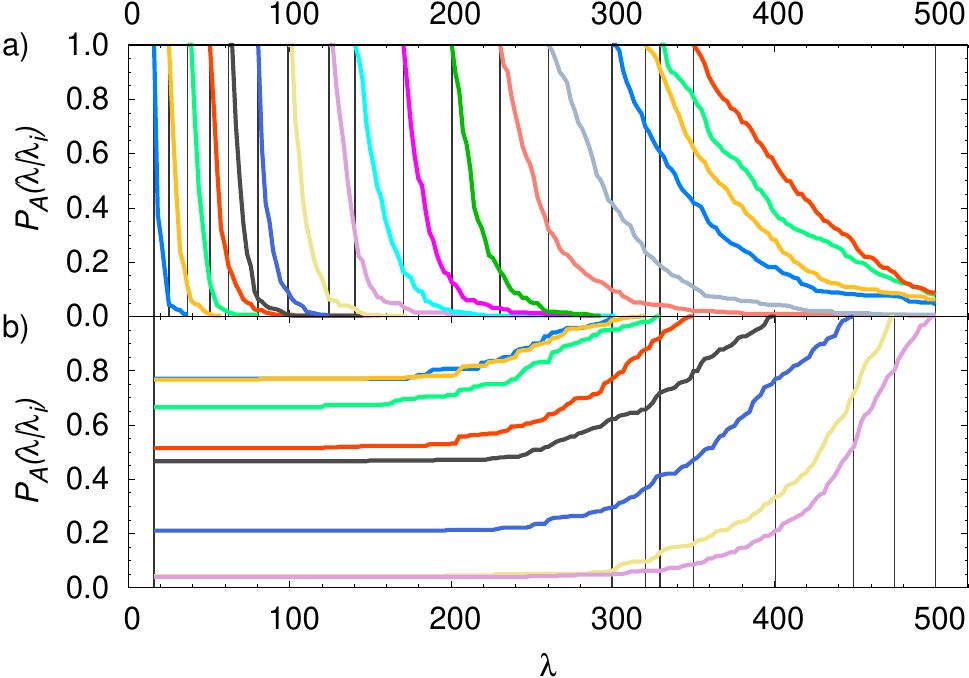}
\caption{Forward (a) and backward (b) crossing histograms for the RETIS simulation of Ni$_3$Al nucleation. Histograms are calculated from 360 paths in each ensemble, spaced five moves apart. Interface locations are marked with grey vertical lines. The rightmost interface in (a) and the leftmost interface in (b) are the boundaries of the solid state and liquid state, respectively.}
\label{fig:ni3al-hist}
\end{figure}

As an example, we discuss results from our previous work on nucleation in Ni$_3$Al.~\cite{Liang2020} For this system, both forward and backward interface ensembles were collected. The resulting crossing histograms are shown in Fig.~\ref{fig:ni3al-hist}, where $\lambda$ corresponds again to the largest cluster size, $n_s$. For the most part, the crossing histograms look fairly well-converged. 
A first indication of sampling issues is that the crossing probabilities of the forward ensemble do not seem to exhibit a plateau region, even for large values of $\lambda > 300$ (Fig.~\ref{fig:ni3al-hist}(a)).
The backward histograms (Fig.~\ref{fig:ni3al-hist}(b)), however, suggest that the critical cluster size should be roughly between 350 and 400---paths in these interface ensembles have an approximately 50\% probability to reach the stable state $A$ (liquid). This estimate of the transition state is not corroborated by the forward crossing histograms, which seem to indicate that cluster sizes up to at least $\lambda = 350$ have a probability of 
less than 10\% to commit to state $B$ ($\lambda \geq 500$, solid basin). Consequently, the crossing histograms are not converged.

\begin{figure}
\centering
\includegraphics[width=0.6\linewidth]{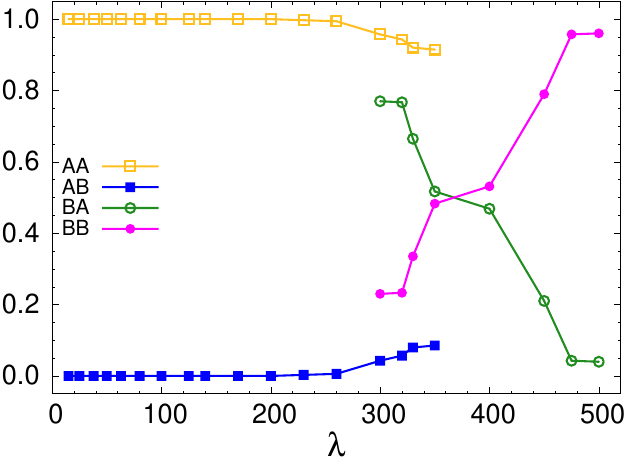}
\caption{Fraction of paths of each type from each interface ensemble collected from the RETIS simulation of Ni$_3$Al nucleation. The fractions of AA and AB paths, as well as the fractions of BB and BA paths, each add to one.}
\label{fig:ni3al-pathtypes}
\end{figure}

A similar conclusion is reached by looking at the number of each type of path in each interface ensemble---i.e., the number of $A$-to-$A$ ($AA$), $AB$, $BA$, and $BB$ paths. The fraction of each path type  is shown in Fig.~\ref{fig:ni3al-pathtypes}. In a converged RETIS simulation, we expect the fraction of each path type to be equivalent to 0.5 at the transition state (here, at the critical nucleus size). However, we again observe that the backward ensembles predict a critical size of $\lambda \approx 350$ (where the fraction of $BB$ and $BA$ paths cross), while in the forward ensembles the critical nucleus size is not yet reached.

Our detailed analysis of the nucleation mechanism in Ni$_3$Al revealed 
that $n_s$ is not a good approximation of the RC.~\cite{Liang2020} 
In this system, multiple channels exist on the free energy landscape that differ not only in the size, but also the crystallinity and chemical short-range order of the critical nuclei. Consequently, trajectories from all possible nucleation pathways need to be included in the path ensemble, which is extremely difficult to sample using only the largest cluster size as the progress coordinate $\lambda$.

Even though the crossing histograms were not converged to the point where a rate constant could be calculated, valuable information about the mechanism was still obtained. In cases such as this, where $\lambda$ is a poor approximation of the RC, an iterative approach to improving $\lambda$ may be needed in order to accurately estimate the rate constant. Such an idea has been applied to FFS,~\cite{Escobedo:07:JCP} where the sampling CV is improved with each successive FFS simulation by fitting a function of multiple CVs to the committor estimates obtained from the path connectivity. The optimal combination of CVs is then used as $\lambda$ to perform another FFS calculation. A similar methodology can be applied to the paths obtained from RETIS. Instead of using the path connectivity, the trajectory outcomes from RETIS simulations can be utilized to estimate the averaged committor and identify the best combination of CVs based on a maximum likelihood estimation approach~\cite{Peters:06:JCP,Rogal2010b}. The optimal CVs can then be used to perform successive rounds of RETIS calculations. While these approaches aid in obtaining better sampling of the TPE and more accurate rate constants, multiple RETIS simulations are naturally computationally expensive.

%---------------------------------------- FFS ----------------------------------------

\subsection{FFS}
\label{subsec:ffs}
Like TIS, efficient sampling in FFS depends on path decorrelation.
All pathways are built upon the initial configurations, hence proper basin sampling is crucial. Proper sampling requires that the basin simulation converges to equilibrium sampling and that the harvested initial configurations are decorrelated. The second important set of parameters are those involved in interface placement. The interfaces cannot be placed too far apart such that only a few successful configurations are obtained, nor too close where the trajectories do not have sufficient time to decorrelate from the configurations they were shot from. Ideally, the pathways obtained at the end of the FFS calculation will not have passed through only a few configurations at any given interface. Such configurations are commonly referred to as bottlenecks. Bottlenecks lead to fewer decorrelated paths, thereby limiting mechanistic analysis and introducing greater uncertainty in the rate estimate. In this section, we provide a practical guide to evaluate the sampling quality of FFS through two illustrative examples: LJ crystal nucleation and XL hydrates nucleation. 

\subsubsection{Computational details}

\paragraph{Basin simulations for LJ FFS.} Basin simulations of the LJ system were initialized by first simulating a liquid at 1.2$T_{m}$ for 50000$\tau$. Beginning at 32000$\tau$,  configurations were recorded every 2000$\tau$, yielding 10 configurations. Each configuration was quenched to 0.78$T_{m}$ and simulated for 5000$\tau$ with a Berendsen thermostat and Berendsen barostat,~\cite{Berendsen:84:JCP} followed by simulation with a CSVR thermostat~\cite{CSVR} and Parrinello--Rahman barostat~\cite{Parrinello-Rahman} for 55000$\tau$, resulting in a total of 10 basin simulations. The last 50000$\tau$ of each of these simulations were used to calculate the flux and harvest initial configurations at $\lambda_0$. In our FFS calculations, $\lambda_A < \lambda_0$ and the flux is calculated by considering only first crossings of $\lambda_0$. First crossings are crossings that more recently crossed $\lambda_A$ rather than $\lambda_0$ when followed backwards in time. Configurations were harvested only if they landed on the interface at the time of first crossing (i.e., $\lambda = \lambda_0$). The probability distribution of $\lambda$ values sampled in the basin simulations was calculated. $\lambda_A$ was placed approximately one half standard deviation above the mean value of $\lambda$. $\lambda_0$ was placed to obtain approximately 1000 decorrelated configurations from the basin simulations. The first crossings harvested at $\lambda_0$ were spaced at least 100$\tau$ apart to ensure adequate decorrelation between them. Further details on assessing the decorrelation are given in Sec.~\ref{sec:ffs-lj}.

\paragraph{Basin simulations for XL FFS.} Initially, a single phase system of XL and mW was equilibrated at 300~K and 1~atm for 1~ns. This was followed by another 25~ns of simulation at 300~K and 500~atm. Beginning at 16~ns, a configuration was saved every 1~ns for total of 10 configurations. From each configuration, ten trajectories were launched at 230~K and 500~atm, with velocities assigned from the Maxwell--Boltzmann distribution. Each of the 100 trajectories was 6~ns in length, and the last 5~ns were used to harvest initial configurations and calculate the flux. $\lambda_A$ and $\lambda_0$ were placed according to similar criteria that were used for the LJ system. A minimum spacing of 500~ps was required between successive harvested crossings.

\paragraph{Interface placement for LJ and XL FFS.} The FFS interfaces were placed on-the-fly. At a given interface, $i$, with $N_i$ configurations, a total of $M_i$ trajectories were initiated. $M_i/N_i$ trajectories were launched from each configuration. The remainder (i.e. $mod(M_i/N_i)$) of trajectories were started from configurations selected randomly without replacement. Ensuring that each configuration has approximately equal number of trajectories launched from it helps reduce genetic drift in the pathways~\cite{Allen2006a,Kunsch:11,Larson:20:Polymers}---that is, the configurations at a given interface have a lower chance of originating from only a few highly reactive configurations at lower interfaces. Every configuration had trajectories initiated from it and hence a possibility of crossing $\lambda_{i+1}$. This reduced the chances that some pathways were terminated prematurely, thereby decreasing the degree of correlation between configurations at the next interface. Each simulation was run for a predetermined length. For the LJ system we used 5$\tau$, and for XL systems this length was 5 ps. After running the trajectories for the trial length, $\lambda_{i+1}$ was placed to obtain the approximate number of desired configurations. The remaining incomplete trajectories were continued until they reached $\lambda_{i+1}$ or the basin. Configurations with $\lambda=\lambda_{i+1}$ were stored to initiate trajectories from $\lambda_{i+1}$. Our interface placement algorithm is similar to the exploring scouts method,\cite{Allen:13:JCP} except that the full set of simulations is launched at the start of a new interface instead of a smaller trial set. 

\paragraph{Collective variables.} Sampling checks for FFS require the calculation of CVs besides the one used to define the interfaces. Two orthogonal CVs were used for the LJ system. $Q_6^{cluster}$ is a measure of crystallinity of the cluster~\cite{Moroni2005} and is based on the Steinhardt $q_6$ CV~\cite{Steinhardt1983}. The largest cluster size based on a stricter definition of solid particles was also used, denoted $n_{LD}$. This is based on the the neighbor-averaged $q_6$ ($\overline{q}_6$) CV developed by Lechner and Dellago.\cite{Dellago:08:JCP} 
For the calculation of $n_{LD}$, particles within 1.36$\sigma$ are considered neighbors and a $\overline{q}_{6m}$ value of 0.36 or greater is required for a particle to be classified as solid.

We used $BC_{planar}$\cite{DeFever:17:JCP} as the CV for the XL FFS calculations. $BC_{planar}$ is based on the CV developed by B\'{a}ez and Clancy.\cite{Clancy:94:NYAS} For a water molecule to be classified as hydrate-like, it must be part of 4--6 5-membered rings. The largest cluster of the hydrate-like water molecules is then identified.
Two additional CVs were also used for further analysis of the XL FFS calculation. The first was based on mutually coordinated guests (MCG),~\cite{Sum:14:JCP}  where the size of the largest cluster of hydrate-like guest molecules was used as the CV. Guest molecules were labeled as hydrate-like based on being mutually coordinated. Two neighboring guest molecules were considered to be a mutually coordinated pair when the overlapping cones projected from the guest molecules had at least five water molecules within them. A guest molecule had to be part of at least two such mutually coordinated pairs to be classified as hydrate-like. The second CV was the dihedral order parameter (DHOP)~\cite{DeFever:17:JCP}, where a water molecule was labeled hydrate-like if it  participated in 11--12 planar dihedrals with its nearest neighbors. The size of the largest cluster of hydrate-like water molecules was used as the CV. The full algorithm to calculate DHOP is available in Ref.~\citenum{DeFever:17:JCP}.

\subsubsection{Lennard-Jones crystal nucleation}
\label{sec:ffs-lj}

\begin{figure}[h]
\centering
\includegraphics[width=0.6\linewidth]{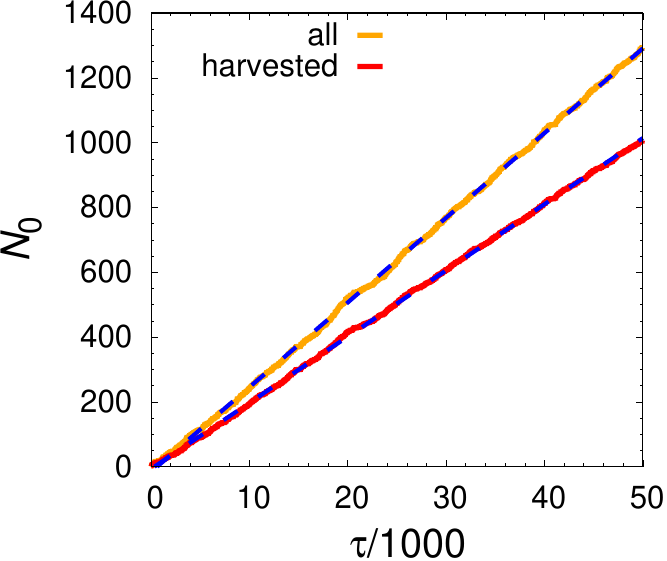}
\caption{Number of total and harvested first crossings of the first interface versus time for the LJ basin simulations. The blue dashed lines are linear fits to the data. The results are summed over 10 basin simulations of 50000$\tau$ each.}
\label{fig:flux-lj}
\end{figure}

The quality of initial configurations collected at $\lambda_0$ is critical to obtaining good sampling in FFS. This quality primarily depends on the extent of the basin sampling and the decorrelation between successive configurations. One of the first checks for convergence of the basin simulation is the number of first crossings and the number of harvested crossings versus basin simulation time. We expect a linear behavior indicating that the flux is constant throughout the simulation. For the LJ system, the number of first crossings and the number of harvested crossings (Fig.~\ref{fig:flux-lj}) versus time are linear. If non-linear behavior is observed, it could indicate a lack of equilibrium sampling and that the system has not relaxed.

The flux through $\lambda_0$ alone is not a sufficient check for convergence of the basin simulation, as was demonstrated by Bi and Li in their study of methane hydrate nucleation with a gas-liquid interface~\cite{Li:14:JPCB}. 
Even though the flux showed a linear behavior with simulation time, the overall crossing probability, $P_{A}(\lambda|\lambda_0)$, showed significant dependence on the length of the basin simulation. This was attributed to a nonequilibrated spatial distribution of the hydrate nuclei relative to the distance from the gas--liquid interface. This suggests that it is important to check for relaxation (and convergence) of the basin simulation along multiple degrees of freedom, especially in heterogeneous systems.

\begin{figure}[h]
\centering
\includegraphics[width=0.6\linewidth]{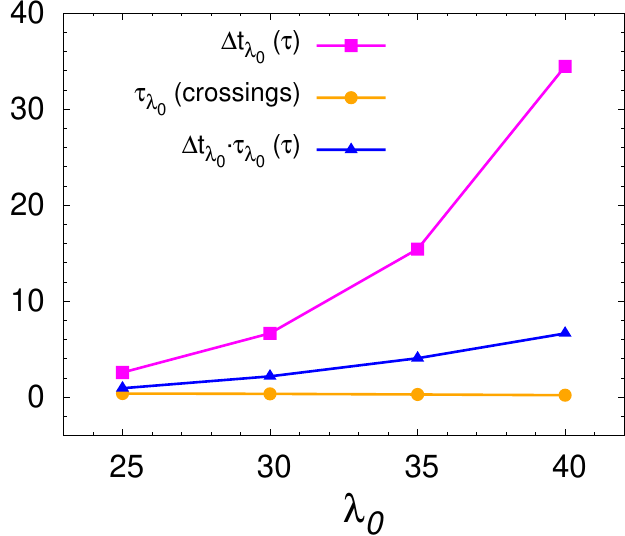}
\caption{Dependence of $\tau_{\lambda_0}$ and $\Delta t_{\lambda_0}$ on the position of $\lambda_0$ for the 200000-$\tau$ LJ basin simulation. Units of the ordinate for each curve are in parentheses in the legend. $\lambda^{\perp}$ is $Q_6^{cluster}$. $\lambda_A$ is located at a value of 21.}
\label{fig:acf-lj}
\end{figure}

It is also essential that the first crossings harvested at $\lambda_0$ are decorrelated. Successive first crossings at $\lambda_0$ can have some degree of correlation even when $\lambda_0 > \lambda_A$. We use the method proposed by Velez-Vega et al.~\cite{Escobedo:09:JCP} to obtain decorrelated initial configurations at $\lambda_0$. In this method, a CV orthogonal to $\lambda$ is calculated for all first crossings for several trial values of $\lambda_0$. This orthogonal CV, $\lambda^{\perp}$, gives a measure of phase space explored in directions distinct from $\lambda$. For nucleation, examples of $\lambda^{\perp}$ include the composition of the nucleus, location of the nucleus relative to an interface (for heterogeneous nucleation), or nucleus size measured using different criteria. The ACF of $\lambda^{\perp}$ versus the lag measured in number of first crossings is calculated as follows:
\begin{equation}
    \mathrm{ACF}(\mathrm{lag})=\sum_{i=1}^{N-\mathrm{lag}} \frac{\left(\lambda^{\perp}_{i}-\overline{\lambda^{\perp}}\right)\left(\lambda^{\perp}_{i+\operatorname{lag}}-\overline{\lambda^{\perp}}\right)}{\sum_{i=1}^{N}\left(\lambda^{\perp}_{i}-\overline{\lambda^{\perp}}\right)^{2}},
\end{equation}
where $\overline{\lambda^{\perp}}$ is the average value of $\lambda^{\perp}$ and $N$ is the number of first crossings. Each ACF is fit to a function proportional to $\exp(-\mathrm{lag}/\tau_{\lambda_0})$ to obtain the autocorrelation time $\tau_{\lambda_0}$. $\tau_{\lambda_0}$ can also be estimated by using Eq.~(\ref{eq:tau-acf}). For each trial $\lambda_0$, the average time between first crossings, $\Delta t_{\lambda_0}$, is also calculated. The minimum required time between decorrelated configurations at $\lambda_0$ is then the product $\tau_{\lambda_0} \times \Delta t_{\lambda_0}$. All three values for the LJ basin simulation are shown in Fig.~\ref{fig:acf-lj}, with $\lambda^{\perp}$ equal to the $Q_6^{cluster}$ value of the largest cluster in the configuration. The goal is to find the minimum value of  $\tau_{\lambda_0} \times \Delta t_{\lambda_0}$, which would in turn maximize the number of decorrelated configurations obtained per unit of simulation time. For this system, we do not see a minimum in $\tau_{\lambda_0} \times \Delta t_{\lambda_0}$. Instead, $\tau_{\lambda_0}$ remains small for all trial $\lambda_0$ values, and the product is dominated by $\Delta t_{\lambda_0}$. We observed similar behavior for two additional definitions of $\lambda^{\perp}$: $n_{LD}$ and the distance of the largest cluster center of mass from the origin of the box. Smaller values of $\tau_{\lambda_0}$ relative to  $\Delta t_{\lambda_0}$ suggest that most of the first crossings at $\lambda_0$ are distinct clusters for all values of $\lambda_0$. It is important to note that we are only considering the first crossings that land directly at $\lambda_0$. If all first crossings are considered for the initial configurations, then the behavior may be different since $\Delta t_{\lambda_0}$ will decrease. The ACF is not the only possible measurement of decorrelation between configurations. In nucleation, for example, one can use the percentage of particles shared by the largest solid cluster in two configurations, and $\tau_{\lambda_0}$ can be selected as the time when the percentage passes below a defined threshold.

\begin{figure}[h]
\centering
\includegraphics[width=0.55\linewidth]{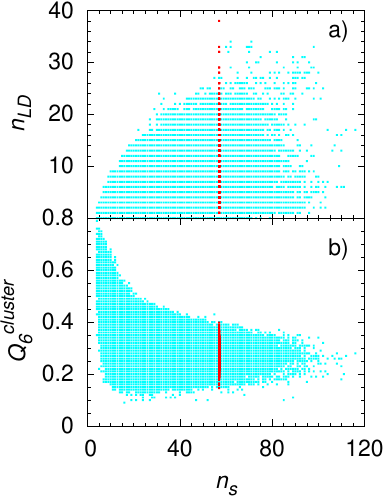}
\caption{Sampling of the LJ basin by a 200000-$\tau$ simulation (cyan points) for (a) $n_{LD}$ and (b) $Q_6^{cluster}$. Red points are the initial configurations harvested at $\lambda_0$ from the original basin simulations.}
\label{fig:lj-basin}
\end{figure}

In addition to having decorrelated configurations at $\lambda_0$, it is important to have extensive sampling of the orthogonal CVs along $\lambda_0$. This is because there is the possibility that the density of configurations at $\lambda_0$ as a function of some orthogonal CV $\lambda^{\perp}$ has little overlap with values of $\lambda^{\perp}$ with high reactivity.\cite{VanErp2012} This is particularly true when the sampling CV $\lambda$ is a poor approximation of the RC. One approach to probe the extent of sampling is to perform exhaustive sampling of the basin by using advanced sampling methods such as replica exchange MD~\cite{Okamoto:99:CPL} or metadynamics~\cite{Laio2002}. The orthogonal CVs sampled at a given initial interface value ($\lambda_0$) in these simulations can be compared with those sampled by the configurations harvested at $\lambda_0$ from the basin simulations. For the LJ system discussed here, we ran longer simulations of the liquid basin to assess the sampling along orthogonal CVs.  A 200000-$\tau$ simulation in the basin was performed to assess the orthogonal sampling of $Q_6^{cluster}$ and $n_{LD}$. Note that the configurations at $\lambda_0$ were harvested from 10 basin simulations of 50000$\tau$ each. Fig.~\ref{fig:lj-basin} shows the sampling of orthogonal CVs $Q_6^{cluster}$ and $n_{LD}$ obtained from the long simulations. The values of $Q_6^{cluster}$ and $n_{LD}$ sampled by the configurations at $\lambda_0$ are overlaid on this distribution. The initial configurations span the basin sampling of both CVs at $\lambda_0$, indicating that the use of 50000-$\tau$ simulations in the original basin simulations did not restrict the sampling in the orthogonal directions.

After the configurations at the initial interface are harvested, the next step is to begin sampling the interfaces. It is recommended to maintain a constant flux of trajectories moving forward at each interface and to sample a diverse set of paths.~\cite{Escobedo:08:JCP} The constant flux requirement is ensured by our interface placement algorithm. The diversity of paths, on the other hand, is not easy to control during the FFS calculation. Thus, it should be carefully assessed during and after the FFS calculation. When talking about diversity of paths, the goal is to have several decorrelated transition paths. The correlation is closely related to the extent to which segments of the trajectories are shared between paths. Thus, the diversity of paths is reduced if many paths go through only a small number of configurations at a given interface, referred to as bottlenecks. This problem is further exacerbated if the bottlenecks appear at later interfaces. While this is a recognized issue,~\cite{Allen2006a} there are few established methods to quantify path diversity. Here, we attempt to provide several different measures to this end, with the objective of identifying these bottlenecks---either on-the-fly or after FFS is complete.

\begin{figure}[h]
\centering
\includegraphics[width=0.8\linewidth]{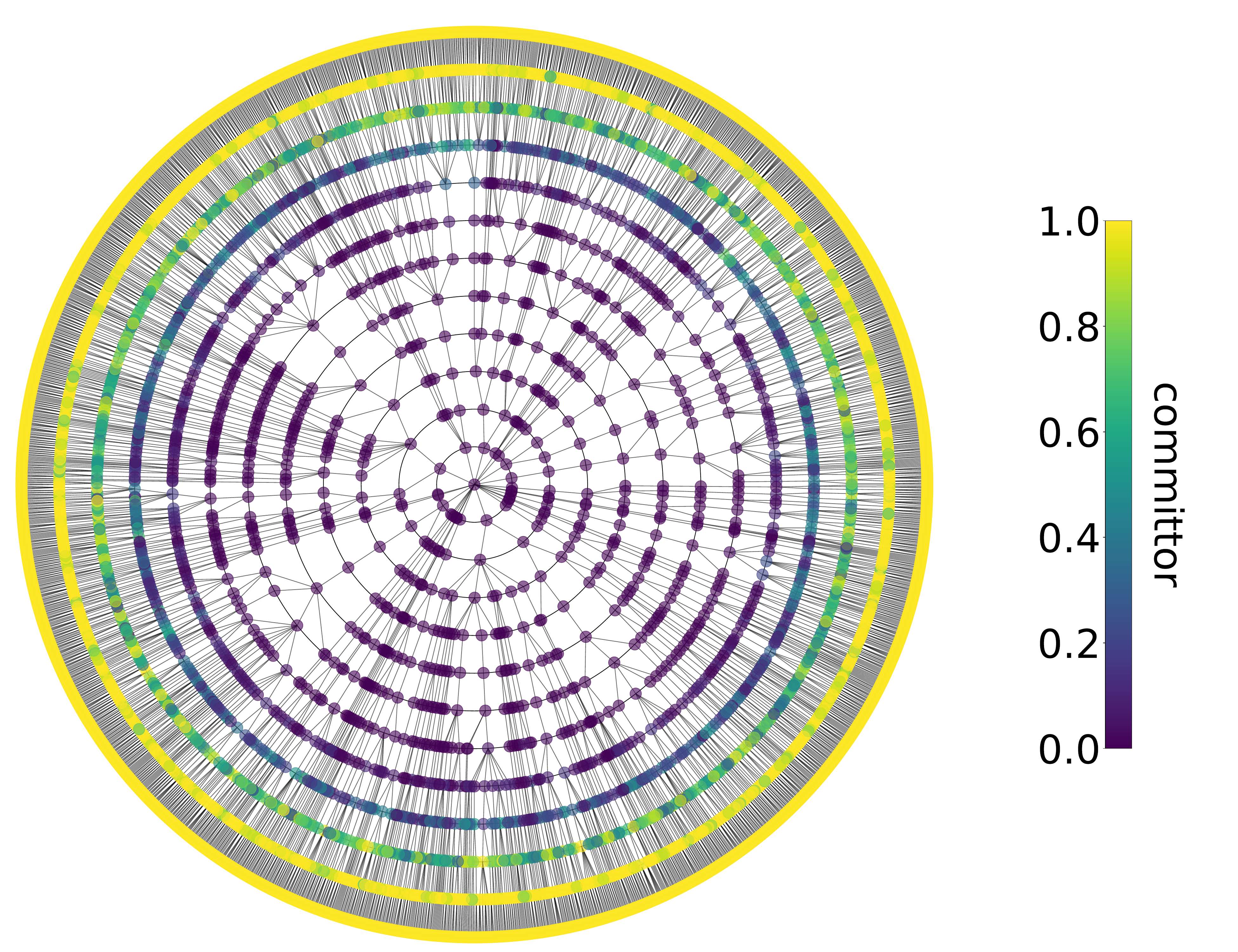}
\caption{Illustration of path connectivity from FFS for the LJ system. The center filled circle represents the basin. Each concentric circle moving away from the center is a subsequent interface (the outermost concentric circle is $\lambda_n=\lambda_B$). Each point on a concentric circle represents a configuration at that interface. Line segments between the points are trajectories. The color represents the committor probability of a configuration and is estimated from the connectivity.\cite{Escobedo:07:JCP}}
\label{fig:tree-lj}
\end{figure}

The first tool to assess the diversity of paths is the connectivity graph. Connectivity graphs provide a visual way to assess whether any obvious bottlenecks are present in the FFS calculation. The connectivity graph traces the ancestry of the last-interface configurations back to $\lambda_0$ while illustrating their connectivity along the transition paths. The connectivity graph for LJ nucleation FFS is shown in Fig.~\ref{fig:tree-lj}. The circle at the center of the graph represents the basin. Each concentric circle moving away from the center is a subsequent interface (the outermost concentric circle is $\lambda_n=\lambda_B$). Each point on a concentric circle represents a configuration at that interface. The points are colored based on their committor probability estimated from FFS~\cite{Escobedo:07:JCP}. The committor of a configuration is the probability that a trajectory initiated from the configuration with randomized velocities will reach the product state before the reactant state. The lines in the connectivity graph represent the trajectories connecting two configurations. Ideally, the transition paths should proceed through many configurations at each interface, which would manifest as high density of points on each circle in the connectivity graph. In case of the LJ system, the connectivity graph (Fig.~\ref{fig:tree-lj}) indicates that the transition paths traverse through a variety of configurations at each interface. In addition, the committor probabilities of the different points at each circle are similar. No configuration that is shared by the majority of transition paths is observed  at any interface, indicating the absence of bottlenecks.

\begin{figure}[h]
\centering
\includegraphics[width=\linewidth]{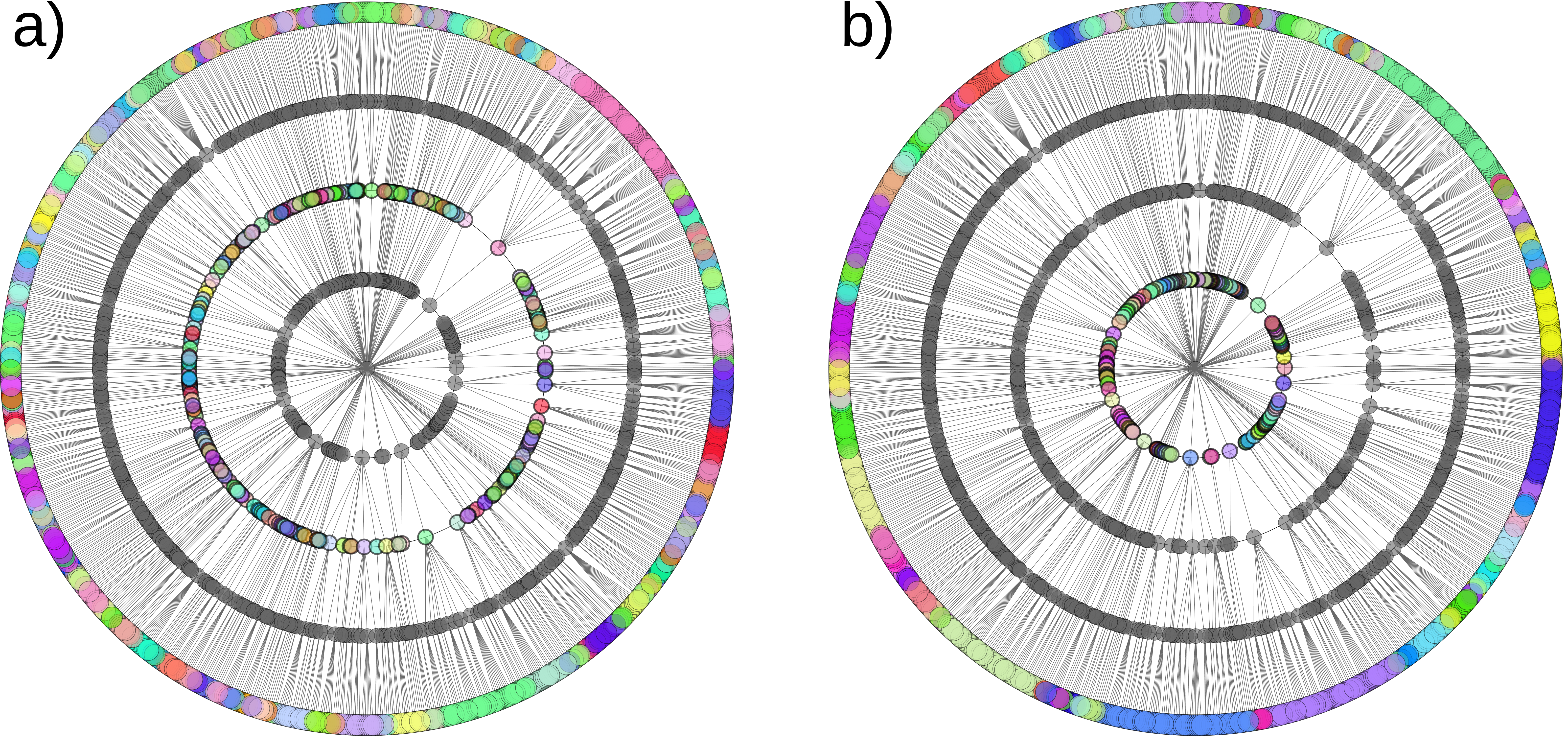}
\caption{Grouping of configurations at interface 3 from the LJ FFS calculation. Configurations were grouped based on common ancestors at interface 1 (a) and 0 (b). Each color represents one group.}
\label{fig:grouping-lj}
\end{figure}

The connectivity graph shown in Fig.~\ref{fig:tree-lj} is informative about the transition path connectivity at the {\it end} of an FFS calculation. However, it is desirable that path diversity can be measured on-the-fly for earlier interfaces to check for early signs of bottlenecks. The approaches we develop here rely on grouping configurations together at a given interface, $i$, based on their common ancestors. For each previous interface $i-n$, where $n \leq i$, different groupings of configurations at interface $i$ are obtained. Such grouping provides insights into the origins of the configurations at a given interface. Ideally, there should be several groups per interface and the size of the groups should be similar. This method of grouping configurations was recently used by Huston et al.~\cite{Larson:20:Polymers} to calculate the correlation between paths and estimate the uncertainty in crossing probabilities. This grouping information can be visualized through connectivity graphs, as shown in Fig.~\ref{fig:grouping-lj}.

In Fig.~\ref{fig:grouping-lj}(a), the grouping of configurations at interface 3 for LJ FFS is shown. Each color on the outer circle is a distinct group of configurations defined by the configuration at interface 1 from which they spawned (so $n=2$). The graph suggests that there are several groups at interface 3 and no single group is dominant. As we trace the ancestry further back to interface 0 ($n=3$), it can be seen in Fig.~\ref{fig:grouping-lj}(b) that several of the groups merge compared to the case where $n=2$. However, there is still no emergence of a dominant group. 

\begin{figure}[h]
\centering
\includegraphics[width=0.95\linewidth]{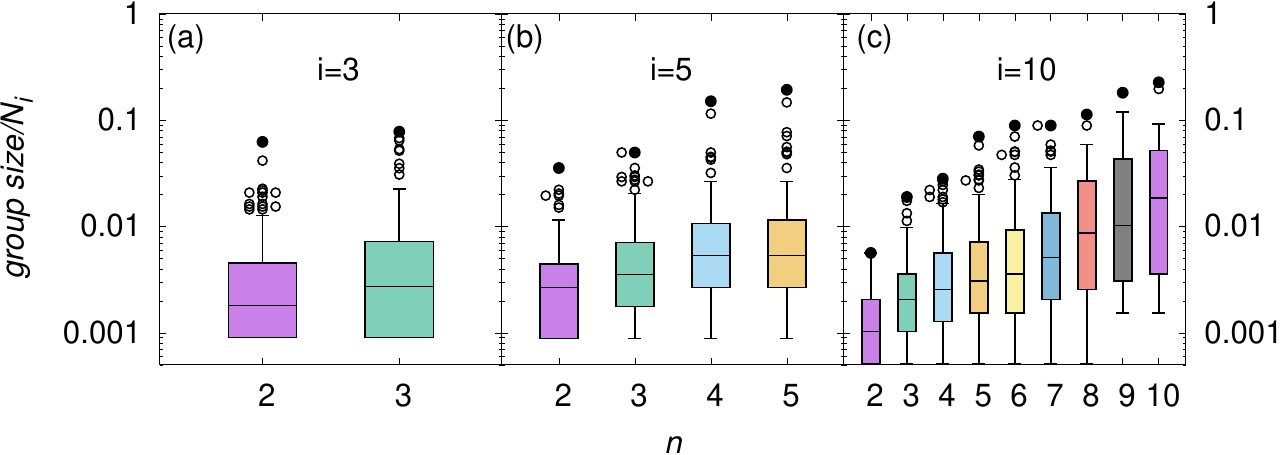}
\caption{Group size distributions for the LJ FFS calculation for (a) interface 3, (b) interface 5, and (c) interface 10, and for every value of $n$ such that $1 < n \leq i$. The group sizes are divided by the total number of configurations at interface $i$. The line in each rectangle is the median and the lower and upper edges of the rectangle are the first and third quartiles. From each of the first and third quartiles, a line is drawn to the data point furthest from the median such that the length of the line from the box edge to the data point is less than or equal to 2.5 times the length of the interquartile range. Group sizes further from the median are represented as circles. The largest group at each interface is marked with a filled circle.}
\label{fig:boxwhisker-lj}
\end{figure}

\begin{figure}[h]
\centering
\includegraphics[width=0.6\linewidth]{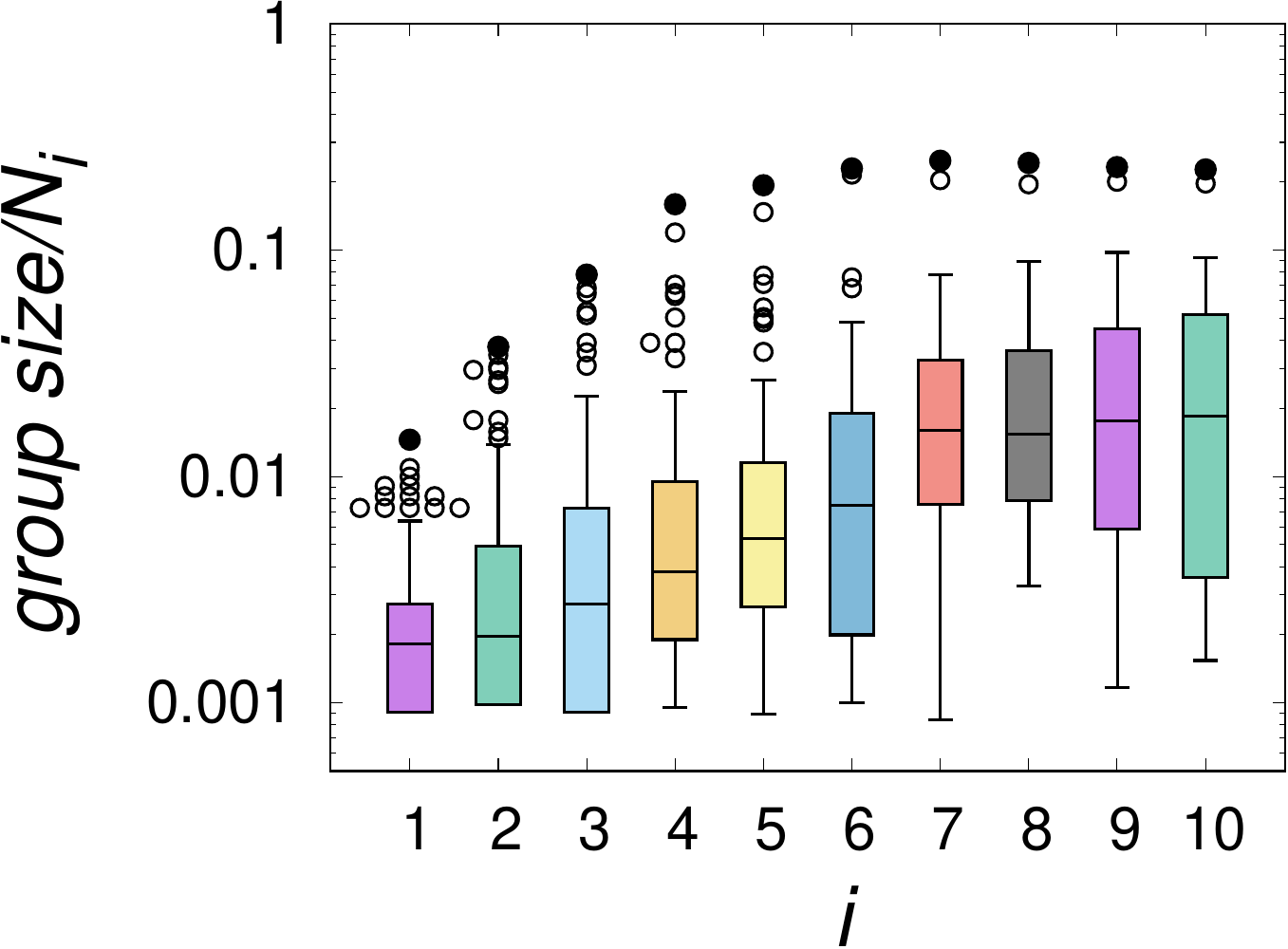}
\caption{Group size distributions for the LJ FFS calculation for each pair of $i$ and $n$ such that $i=n$. That is, the groups at interface $i$ are based on their ancestry at $i=0$. The group sizes are divided by the total number of configurations at interface $i$.}
\label{fig:groupdist-lj}
\end{figure}

A related measure is the group size distribution, which can also be calculated using the connectivity of all paths going from $i-n$ to $i$. This provides information on the extent to which each configuration at $i-n$ contributes to interface $i$. For each interface $i$, groups are constructed for the range of values $n=\{1,i\}$, which assesses the groups at interface $i$ based on common ancestors from each interface below $i$ (i.e., $i-n$). In the case where $n=1$, we obtain information equivalent to the origin histograms used by Dittmar and Kusalik.~\cite{Kusalik:16:JCP} Examples of group size distributions for interfaces 3, 5, and 10 are shown in Fig.~\ref{fig:boxwhisker-lj}. When $i=3$, $n=2$ and $n=3$ correspond to groups of configurations that are spawned from a common ancestor on interface $1$ and interface $0$, respectively (Fig.~\ref{fig:boxwhisker-lj}(a)). 
The distribution of group sizes is normalized by the number of configurations at the interface. In the plots, the upper and lower edges of each rectangle indicate the first and third quartiles, and the line inside the rectangle represents the median. For each of the first and third quartiles (box edges), a line is drawn to the data point furthest from the median such that the length of the line from the box edge to the data point is less than or equal to 2.5 times the length of the interquartile range. The group sizes that are beyond these bounds are shown as points, with the largest group of each interface marked with a filled circle. For the FFS calculations of LJ nucleation, we observe that most of the groups contain less than 10\% of the total configurations for all interfaces. Each interface has a few larger groups but the maximum group size we observe contains less than $20\%$ of the configurations. At interface 10 (Fig.~\ref{fig:boxwhisker-lj}(c)), the group sizes gradually increase as the ancestor configurations move towards inner interfaces (i.e., towards $\lambda_0$). In Fig.~\ref{fig:groupdist-lj} we focus on the group size distributions for every pair of $i$ and $n$, where $i=n$. This corresponds to grouping configurations at interface $i$ by their ancestors at $i=0$. As $i$ increases, the median and maximum group sizes comprise larger fractions of the configurations at each interface. At most, about 30\% of configurations at any interface come from a single configuration at $i=0$. It reaches the maximum around interface 6, but does not increase with subsequent interfaces. No dramatic increase of the group sizes and no appearance of one dominant group is observed. Collectively, these results indicate that there are no bottlenecks in the sampling and signal towards diversity of transition paths. In the worst-case scenario, every trajectory launched from a configuration and its descendants will result in a first crossing of the next interface and all subsequent interfaces. Under our sampling scheme, $N_i$, the number of configurations at interface $i$, and $M_i$, the number of trajectories launched from interface $i$, are chosen constants. If all trajectories launched from a given configuration result in first crossings of $i+1$, then the number of descendant configurations is $M_i/N_i$. If this continues, then the maximum number of descendant configurations at interface $i+k$ will be $\min\{(M_i/N_i)^k, N_i\}$.

\begin{figure}[h]
\centering
\includegraphics[width=0.6\linewidth]{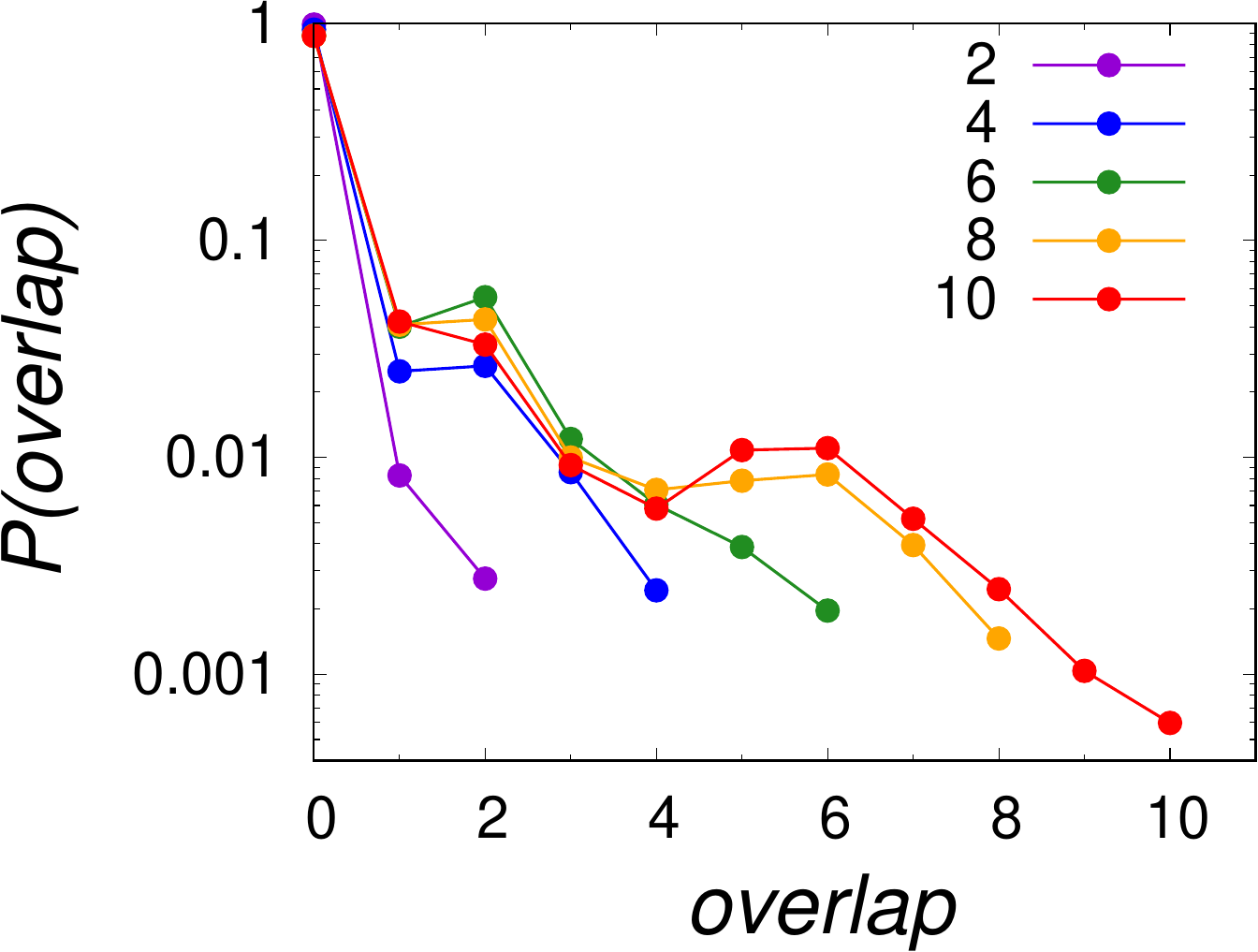}
\caption{Histogram of overlapping trajectory segments across all pairs of paths that reach each interface. The value of the overlap is the number of configurations shared by a pair of paths.}
\label{fig:overlapHist-lj}
\end{figure}

Does the appearance of a bottleneck automatically imply lack of diversity of paths? It can be argued that if the bottleneck occurs at an early interface the paths may be able to sample a diverse path ensemble as they progress through the later interfaces. While this question is challenging to answer, we have developed a measure that can provide some quantitative insights into the correlation between the paths obtained from FFS. The overlap measures the number of configurations shared between two paths. For two paths that go from $\lambda_{0}$ to two configurations $v$ and $w$ at $\lambda_{i}$, the overlap is calculated as
\begin{equation}
    \mathrm{overlap}_{v,w} = \sum_{k=0}^{i-1}{\delta_{\psi_{k,v}\psi_{k,w}}}.
\end{equation}
The sum is calculated over all interfaces prior to $i$. $\psi_{k,v}$ is the index of the configuration at interface $k$ that is an ancestor of configuration $v$. $\delta$ is the Kronecker delta. It is expected that when paths share few configurations they are less correlated with each other. The histograms of overlap values is plotted in Fig.~\ref{fig:overlapHist-lj} for several interfaces of the LJ system. The histograms are calculated over all pairs of paths that reach a given interface. The histograms are peaked at a value of zero for all interfaces, which is expected in the ideal case. The probability of higher overlap (e.g. $>$1 configuration) decreases significantly to less than $0.5\%$. This further reinforces that there are no bottlenecks in our sampling and the paths are diverse.

We used the method similar to that introduced by Huston et al.~\cite{Larson:20:Polymers} to quantify the correlation between paths and estimate the uncertainty in the crossing probabilities. To apply the method, the notion of groups of configurations at each interface as described above was applied. We then determined the number of independent groups at each interface by calculating the intraclass correlation (ICC, or $r_i^{(n)}$) for each $(i,n)$ pair. The ICC quantifies the contribution of group properties to the overall variance of samples. For our FFS calculations, we used the ICC based on one-way random effects analysis of variance~\cite{Halverson:98:JOM}. The calculation details of $r_i^{(n)}$ are provided in the appendix.

\begin{figure}[h]
\centering
\includegraphics[width=0.6\linewidth]{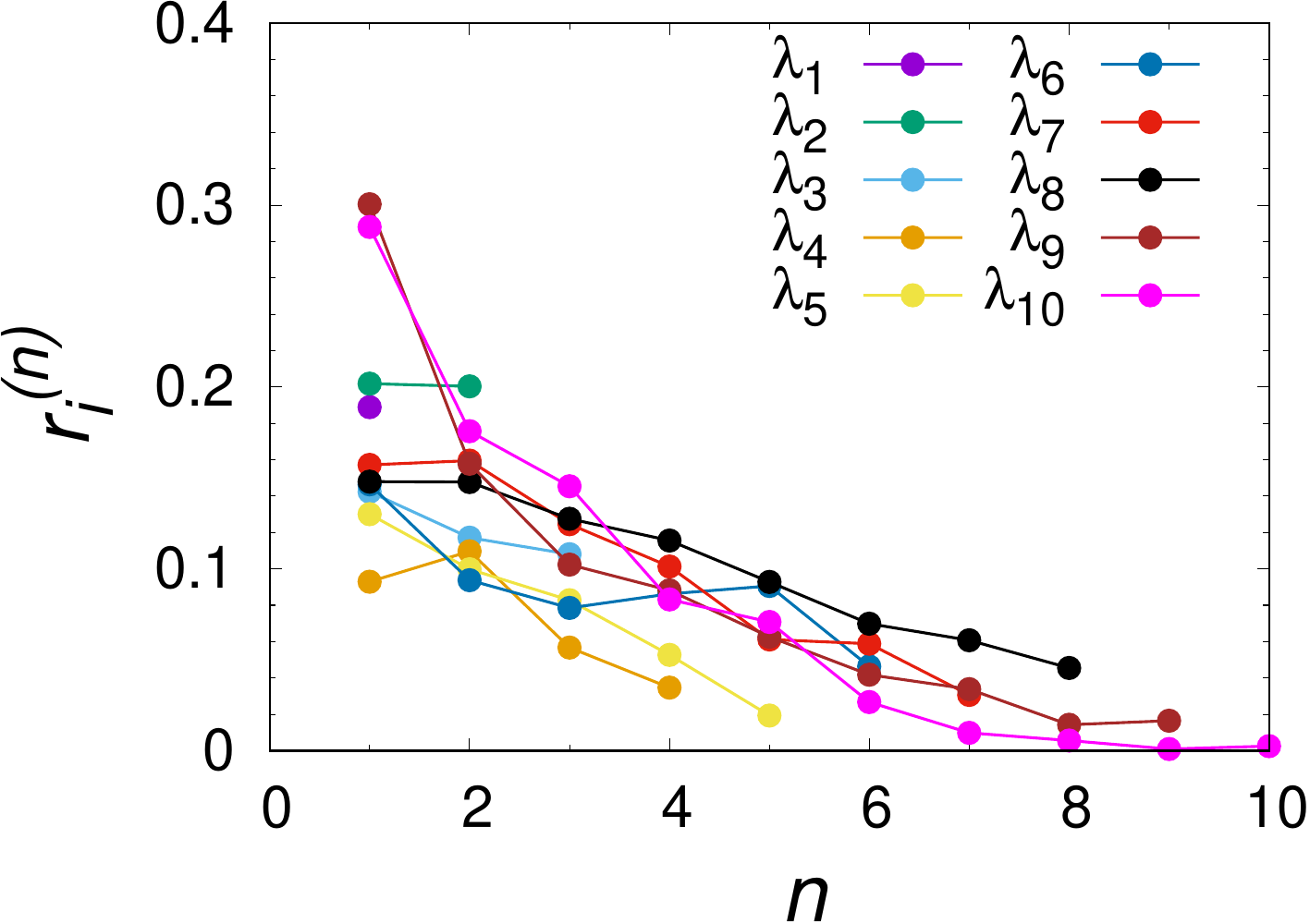}
\caption{Path correlation for the LJ FFS paths calculated using the 
ICC of crossing probabilities. Each curve corresponds to an interface.}
\label{fig:pathcorrelation-lj}
\end{figure}

The resulting $r_i^{(n)}$ values are shown in Fig.~\ref{fig:pathcorrelation-lj}. For each $i$, the smallest possible value of $n$, denoted $L$, was chosen such that $r_i^{(n)} < 1/e (=0.368)$, though the cutoff value can be modified. For all interfaces of the LJ system, $r_i^{(n)}$ is below $1/e$ at $n=1$, so the number of independent groups was found by going back only one interface for every $i$ (so $L=1$). These groups were then used to calculate the standard error of the mean of the crossing probabilities. Further details are provided in the appendix.

\begin{figure}[h]
\centering
\includegraphics[width=0.6\linewidth]{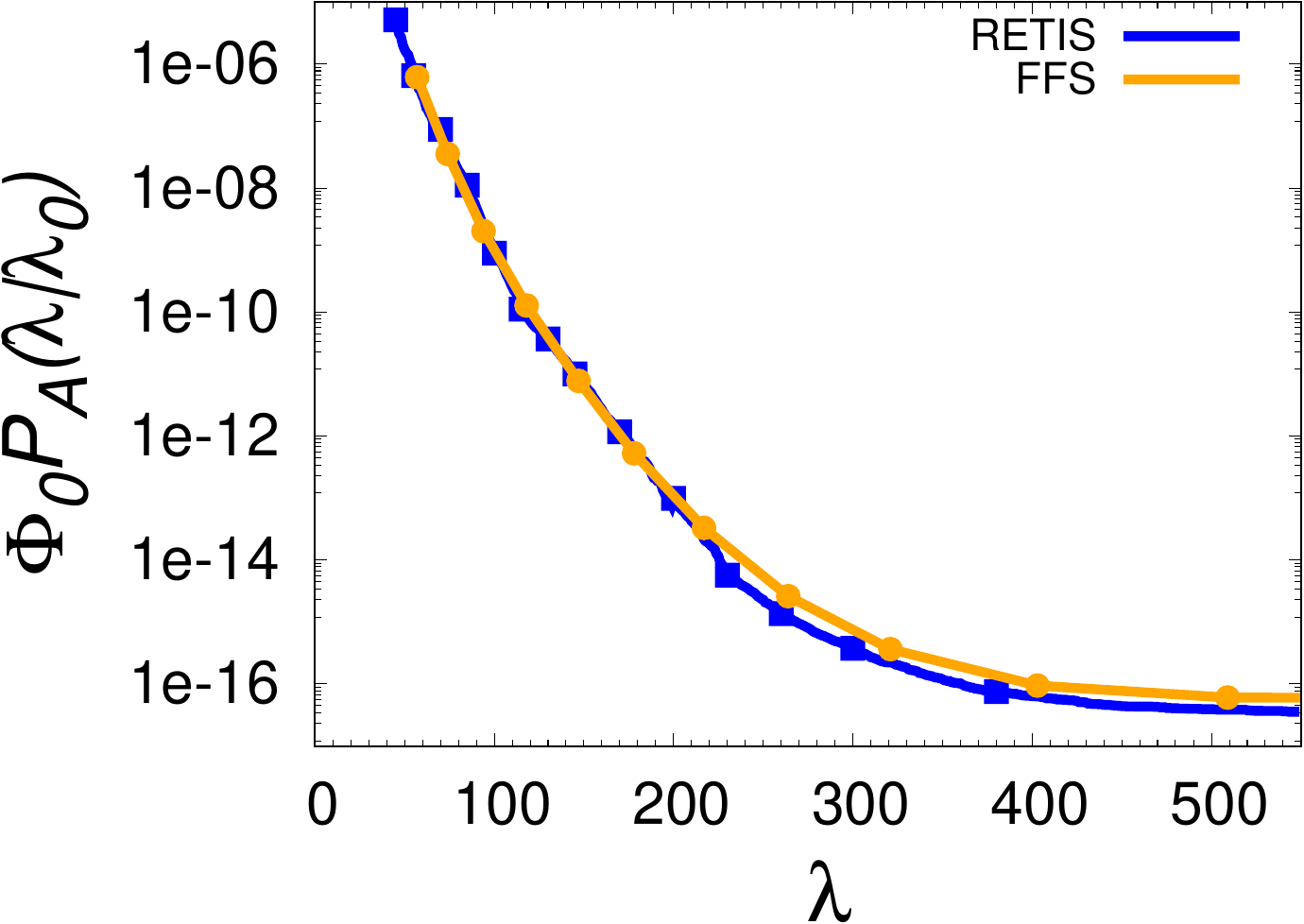}
\caption{Flux through each value of $\lambda$ for rMBV RETIS and FFS. Each point is placed at an interface location.}
\label{fig:retis-ffs-comparison}
\end{figure}

The rate of nucleation obtained from these calculations was $\log_{10}(J_{\text{FFS}}) = -16.2 \pm 0.44$, where the uncertainty was calculated with a 95\% confidence interval. Nearly all of the uncertainty comes from the crossing probabilities. This is in agreement with that obtained from the rMBV RETIS simulations ($\log_{10}(J_{\text{RETIS}})=-16.6 \pm 0.51$). In addition, we see that the flux versus $\lambda$, $\Phi_0 P_{A}(\lambda | \lambda_0)$, between rMBV RETIS and FFS (shown in Fig.~\ref{fig:retis-ffs-comparison}) are in good agreement.

%-------------------- XL hydrate ---------------------------
\subsubsection{XL hydrates nucleation}
\label{sec:ffs-xl}

\begin{figure}[h]
\centering
\includegraphics[width=0.6\linewidth]{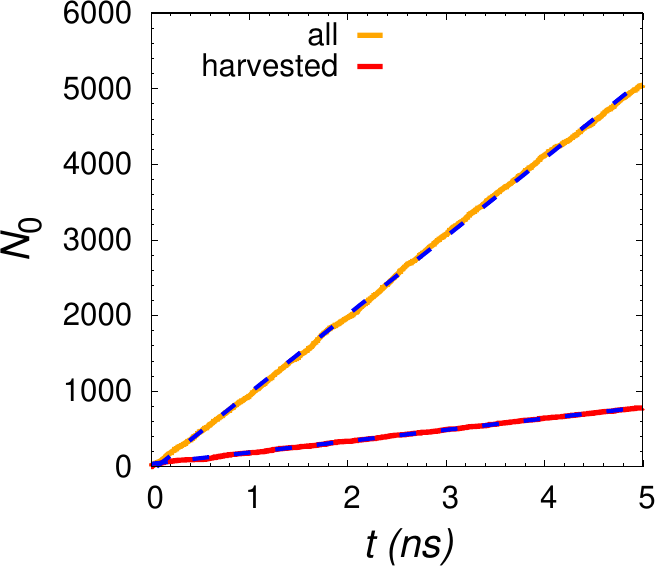}
\caption{Number of first crossings and harvested first crossings versus time for the XL basin simulations. The blue dashed lines are a linear fits to the data. The results are summed over 100 simulations of 5 ns each.}
\label{fig:flux-xl}
\end{figure}

To further demonstrate the practices described in the LJ system to assess the sampling quality  of FFS simulations, we now discuss a more complex system -- nucleation of XL hydrate from a homogeneous two-component liquid. We follow a similar sequence of checks, as described earlier, starting from the basin simulations. As an initial convergence check of the basin sampling, the number of first crossings and harvested configurations over the length of the basin simulations is shown in Fig.~\ref{fig:flux-xl}. The relationship is linear, which is necessary---though not sufficient---for convergence.

\begin{figure}[h]
\centering
\includegraphics[width=0.6\linewidth]{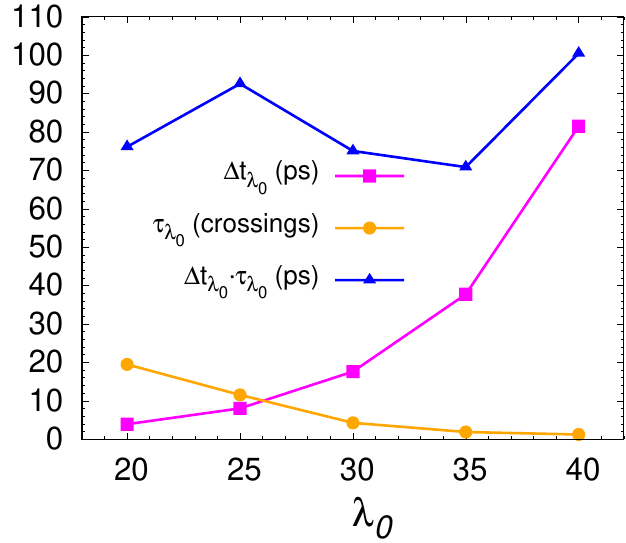}
\caption{Dependence of $\tau_{\lambda_0}$ and $\Delta t_{\lambda_0}$ on the position of $\lambda_0$ for the 100-ns XL basin simulation. Units of the ordinate for each curve are in parentheses in the legend. $\lambda^{\perp}$ is MCG. $\lambda_A$ is located at a value of 19.}
\label{fig:acf-xl}
\end{figure}

We evaluated the decorrelation of initial-interface configurations for the XL basin simulations using  MCG as $\lambda^{\perp}$. The autocorrelation time, $\tau_{\lambda_0}$ of $\lambda^{\perp}$ as a function of different $\lambda_0$ values is shown in Fig.~\ref{fig:acf-xl}. Also shown are average times between first crossings ($\Delta t_{\lambda_0}$) and minimum time required between decorrelated configurations ($\tau_{\lambda_0}\times\Delta t_{\lambda_0}$).  We see that a minimum in $\tau_{\lambda_0}\times\Delta t_{\lambda_0}$ is attained around $\lambda_0=35$, where approximately 70 ps of simulation is needed to obtain decorrelated first crossings. With this spacing, the average value of the ACF between two harvested crossings should be $1/e$, assuming the ACF fits an exponential decay. To further decrease the correlation between successive harvested configurations, the product (70 ps) must be multiplied by a constant $m$, which is the multiple of $\tau_{\lambda_0}$ required to classify configurations as decorrelated and is chosen based on the desired degree of decorrelation. If $m=2.3$, then the ACF will be approximately 0.1 between harvested configurations that are spaced by $m\tau_{\lambda_0}\Delta t_{\lambda_0}$ (since $\exp(-2.3)=0.1$). By multiplying 70 ps by 2.3, we now obtain a required sampling time of about 161 ps between configurations to get an ACF value of 0.1. Regardless of the selected position of $\lambda_0$, decorrelation between successive harvested configurations should be ensured. For the FFS calculation performed here, $\lambda_0$ was chosen to be 41. With $m=2.3$, the required spacing between initial configurations is then approximately 230 ps (since $\tau_{\lambda_0}\Delta t_{\lambda_0}=100$ ps for $\lambda_0=41$), which is lower than the spacing we used (500 ps). Our initial configurations are therefore sufficiently decorrelated.

\begin{figure}[h]
\centering
\includegraphics[width=0.6\linewidth]{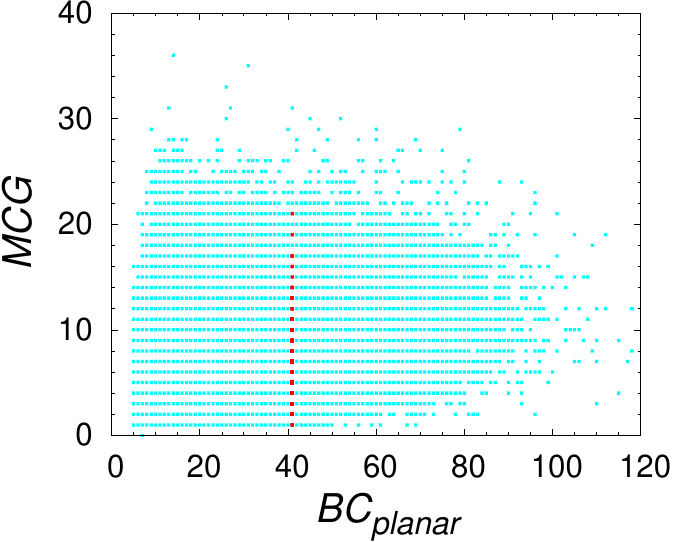}
\caption{Sampling of the XL basin by several long simulations (cyan points). Five simulations were performed for 150 ns each. Due to nucleation events, 100, 50, 20, 20, and 10 ns were used from each of the trajectories. Red points are the initial configurations gathered at $\lambda_0$ from the original basin simulations.}
\label{fig:xl-mcg}
\end{figure}

\begin{figure}[h]
\centering
\includegraphics[width=0.8\linewidth]{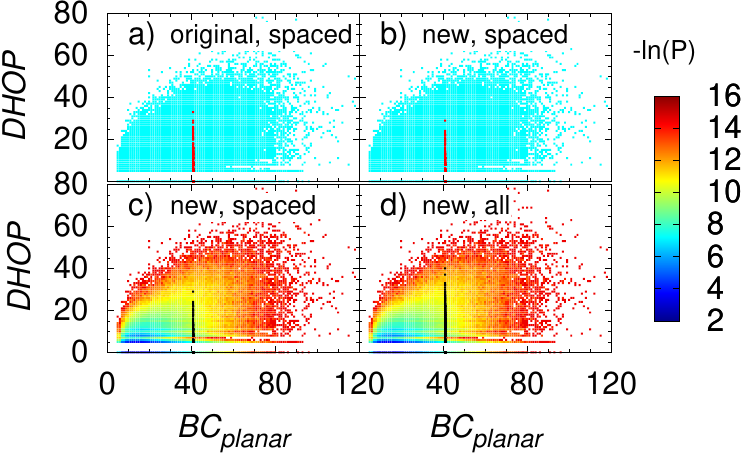}
\caption{Sampling of the orthogonal CV, largest DHOP cluster. 
The cyan and multi-colored points are calculated using the same long basin simulations described in Fig.~\ref{fig:xl-mcg}. Labels within the plots denote the the simulations that the red points in (a) and (b)---and black points in (c) and (d)---were collected from, as well as whether or not the first crossings were spaced by 500 ps. Red points in (a) and (b) are the initial configurations harvested at $\lambda_0$ from (a) the original basin simulations and (b) the new basin simulations, both requiring a minimum spacing of 500 ps between consecutive harvests. Cyan points are all of the pairs of (BC$_{\mathrm{planar}}$,DHOP) values sampled in the new basin simulations. Black points in (c) and (d) are the first crossings from the new basin simulations when (c) first crossings are required to be at least 500 ps apart and when (d) all first crossings are considered.}
\label{fig:xl-dhop}
\end{figure}

In addition to being decorrelated, it is also important that the configurations at $\lambda_0$ represent the distribution of the orthogonal CVs in the basin. To do so, extensive sampling of the basin is required. The original XL basin simulations consisted of 100 5-ns simulations with frames recorded every 0.1~ps for a total of $5\times 10^{6}$ frames. To achieve extensive basin sampling and assess the sampling of the orthogonal CVs along $\lambda_0$, we performed five 150-ns simulations with the same recording frequency. Four of the five simulations nucleated within this time. Thus, only the pre-nucleation part of each trajectory was used to assess the orthogonal sampling (100, 20, 20, 10, and 50 ns were used from each simulation). Sampling results with MCG  and DHOP as the orthogonal CVs are shown in Fig.~\ref{fig:xl-mcg} and Fig.~\ref{fig:xl-dhop}, respectively. With MCG as the orthogonal CV, the initial configurations from the original basin runs span the sampling of the new basin runs. On the other hand, the sampling of initial configurations in the DHOP space appear to be hindered relative to the longer simulations (Fig.~\ref{fig:xl-dhop}(a)). To evaluate whether this is due to the approach used for sampling the basin, we also harvested configurations at $\lambda_0$ from the new basin runs, shown in Fig.~\ref{fig:xl-dhop}(b). No significant improvement in the sampling of the initial interface is observed. While high values of DHOP at $\lambda_0$ are sampled within the timescales of the basin runs, they are generally not harvested as first crossings from the basin. To probe this further, we calculated the probability density of DHOP from the new basin simulations, shown in Fig.~\ref{fig:xl-dhop}(c-d). Also shown in the figures is the sampling of the initial configurations obtained at $\lambda_0$ from the new basin simulations with and without the 500 ps spacing. We see that the regions of low density are sampled by all the first crossings of $\lambda_0$, but enforcing a spacing of 500 ps between first crossings prevents most of the low-density, high-DHOP first crossings from being harvested. This might be addressed by reducing the spacing required between the first crossings (but not below 230 ps) or by running even longer simulations. However, since we already observed nucleation in our 150 ns new basin simulations, the latter may not be a good option. Also, since the high DHOP values are highly unlikely, the former solution  might not resolve the issue. This indicates that our sampling of the orthogonal space at the first interface is reasonable.

\begin{figure}[h]
\centering
\includegraphics[width=0.8\linewidth]{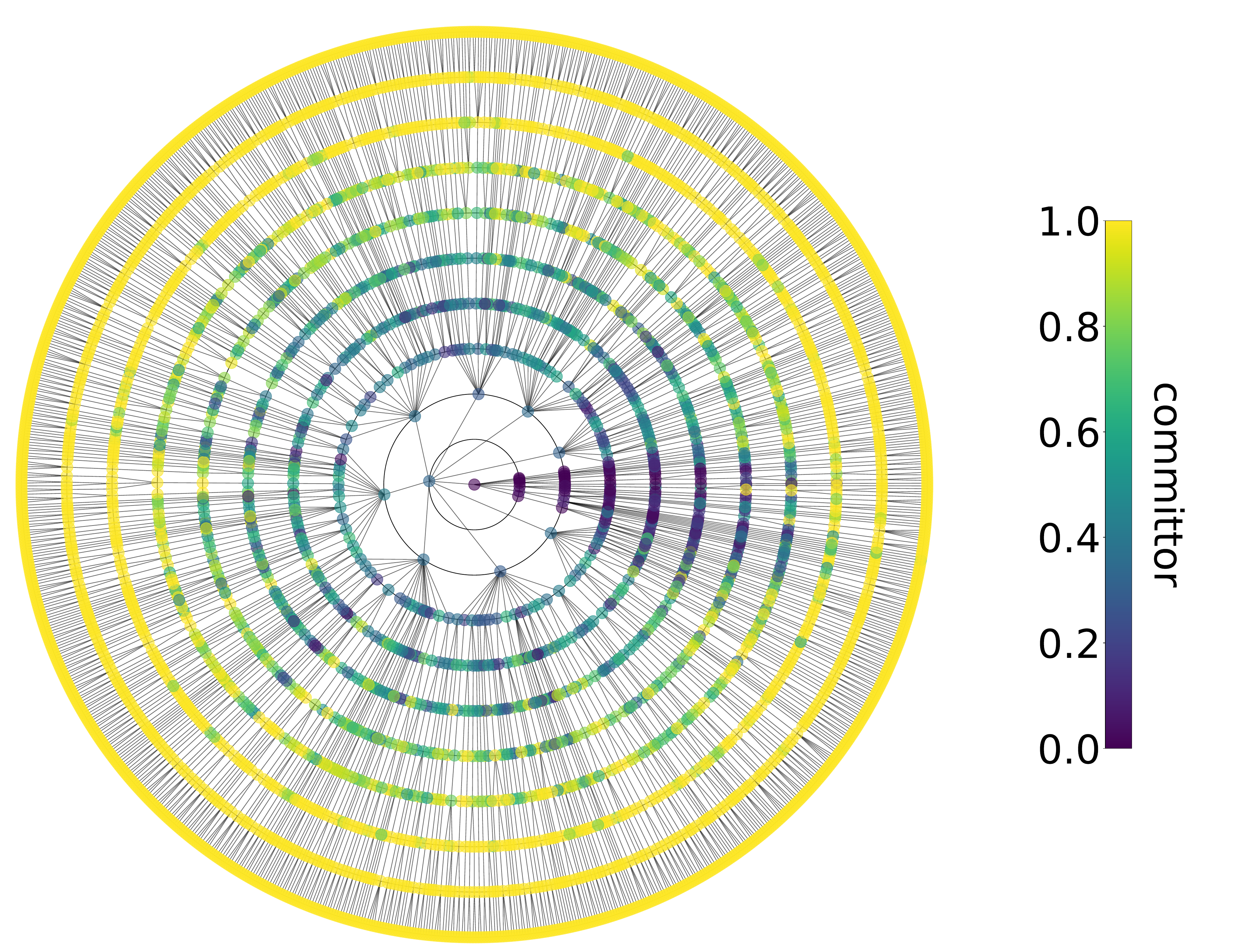}
\caption{Connectivity graph of configurations harvested from the FFS calculation for XL hydrate nucleation.}
\label{fig:tree-xl}
\end{figure}

After ensuring good sampling of the basin and the initial-interface configurations, the next set of assessments focuses on the diversity of paths.
The transition path connectivity graph is presented in Fig.~\ref{fig:tree-xl}. We immediately see that the majority of transition paths ($\sim93$\%) originate from a single configuration at $\lambda_0$. This is intriguing given that we had ensured good sampling of the basin and initial interface. We find that the highly reactive configuration at $\lambda_0$ is near the transition region, based on the committor probability estimate. We surmise that this occurs due to the combination of the conditions of the simulation, where nucleation can be observed on the nanosecond timescale, and a sub-optimal CV. The transition region is accessible on the timescale of the basin simulations, and the CV cannot sufficiently differentiate the transition state configuration from the other initial-interface configurations. This further highlights the challenges associated with rare event sampling in complex energy landscapes. In our original calculations,\cite{DeFever:17:JCP} the bottleneck at $\lambda_0$ was discovered after the completion of the FFS simulations through the connectivity graph. This prompted us to develop methods that could help us detect bottlenecks on-the-fly, i.e., as the FFS simulations are running. The methods described in the previous section (Sec.~\ref{sec:ffs-lj}) are a result of this quest. In the following we discuss the application of these methods to the XL hydrate system.

\begin{figure}[h]
\centering
\includegraphics[width=0.6\linewidth]{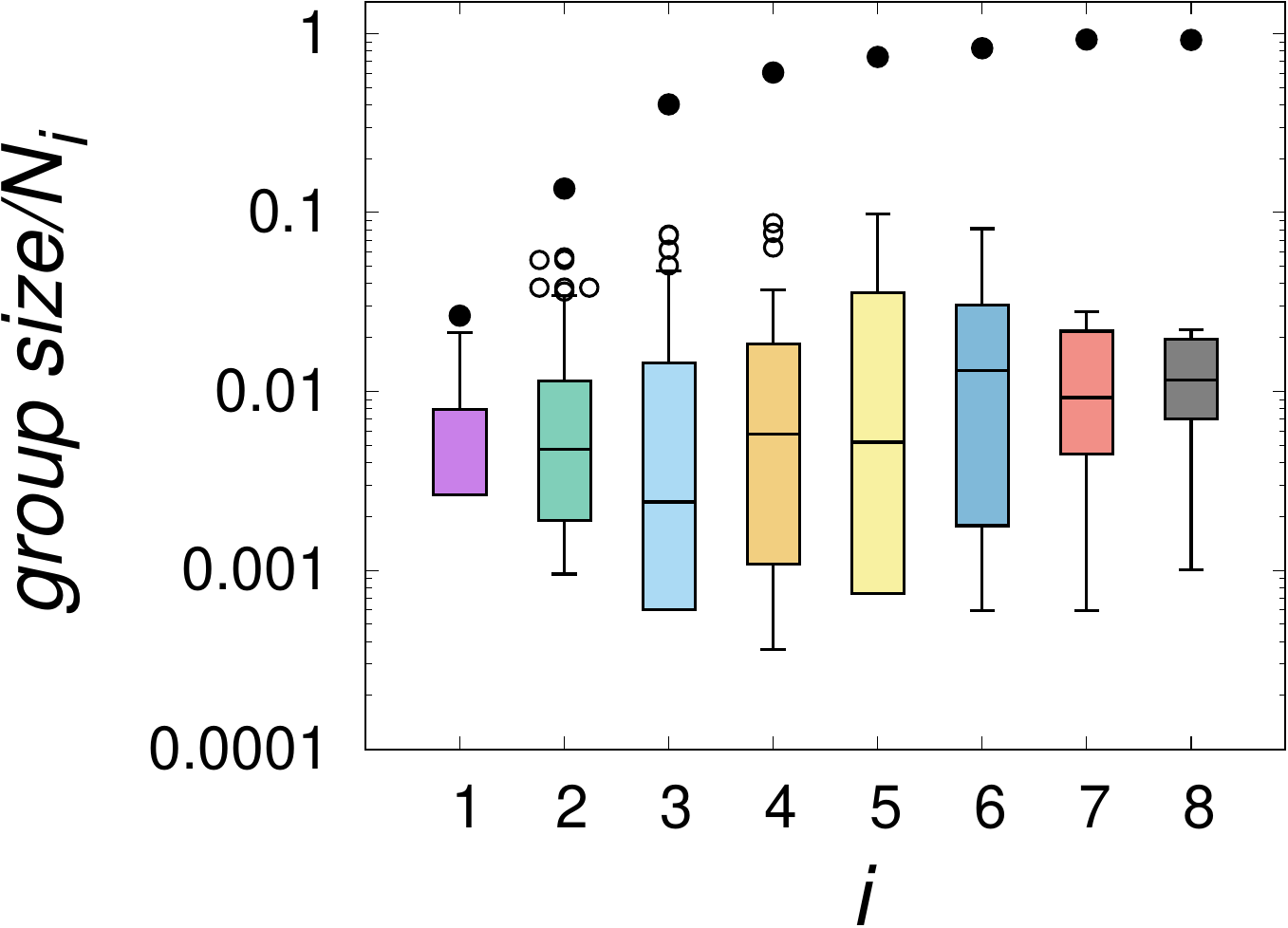}
\caption{Group size distributions for the XL FFS calculation for each $i$ and $n$ pair such that $i=n$. Please refer to the caption of Fig.~\ref{fig:boxwhisker-lj} for a description of the box plot construction.}
\label{fig:groupdist-xl}
\end{figure}

Fig.~\ref{fig:groupdist-xl} shows the group size distributions calculated for the XL hydrate FFS simulations. As described previously, the groups at each interface are defined by the configurations' ancestors at interface 0. Two features emerge:({\it i}) a single group with high percentage of the configurations appears (filled circle at each $i$), and ({\it ii}) this group monotonically increases in size with interface. For example, at interface 3, the largest group has nearly 40\% of the configurations. In other words, 40\% of the configurations at interface 3 are spawned from a single configuration at interface 0. By the final interface, this fraction increases to about 93\%, indicating the presence of a bottleneck at $\lambda_0$. This indicates that using this approach would allow us to identify the bottleneck at interface 0 within three to four interfaces. We recommend making such a plot for each interface and ancestor pair to find bottlenecks at any stage (see Fig.~\ref{fig:boxwhisker-lj}).

\begin{figure*}
\centering
\includegraphics[width=0.8\linewidth]{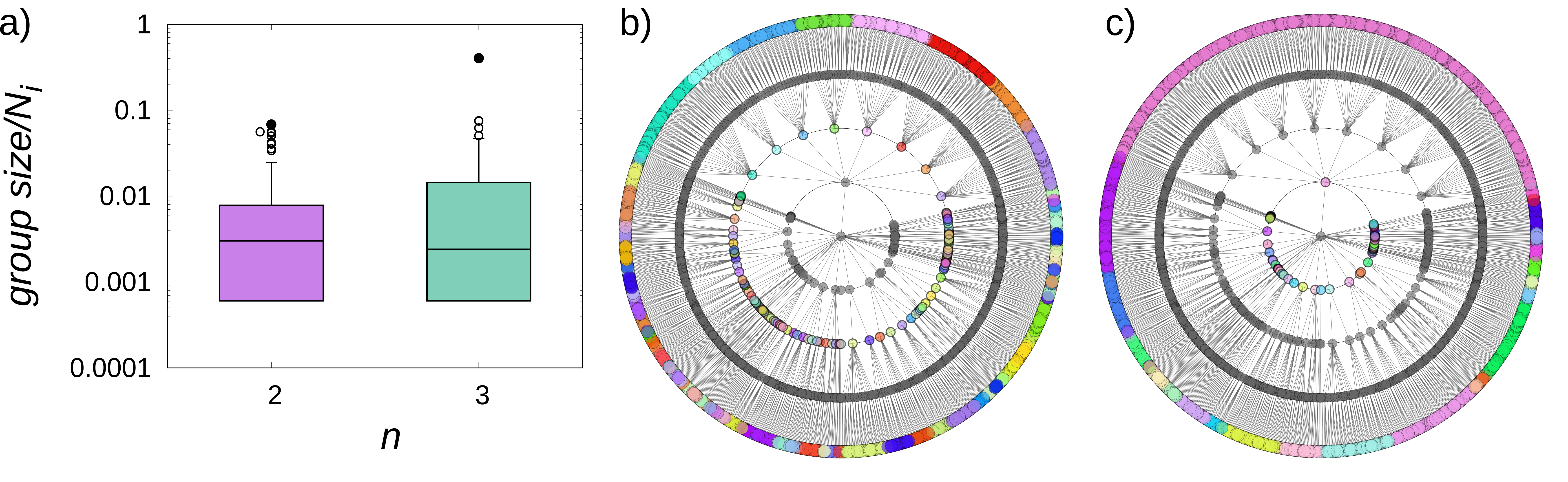}
\caption{(a) Group size distributions for configurations at interface 3 of the XL FFS calculation. For each $n$ shown, the configurations are grouped based on common ancestry back to interface $3-n$. The largest group for each $n$ is shown as a filled circle. (b) and (c) are connectivity graphs that show the groupings of configurations at interface 3 for $n=2$ and $n=3$, respectively. Each color is a distinct group.}
\label{fig:grouping-xl}
\end{figure*}

Fig.~\ref{fig:grouping-xl}(a) shows the group size distributions for $i=3$ with $n=2$ and $n=3$. For $n=2$, the largest group comprises less than 10\% of the configurations. When $n=3$ (meaning the grouping is done based on interface 0) this value jumps to about 40\%. This jump is easily seen in Fig.~\ref{fig:grouping-xl}(b-c). In these connectivity plots, the configurations at a given interface are colored based on the group they belong to. It can be seen that several of the groups merge as $n$ increases for $i=3$. All of these groups originate from the highly-reactive configuration at $i=0$, and at later interfaces, its progeny make up larger portions of the configurations. This indicates that monitoring the group sizes as done on in Figs. \ref{fig:groupdist-xl} and \ref{fig:grouping-xl} provides diagnostics for early signs of bottlenecks in FFS.

\begin{figure}[h]
\centering
\includegraphics[width=0.6\linewidth]{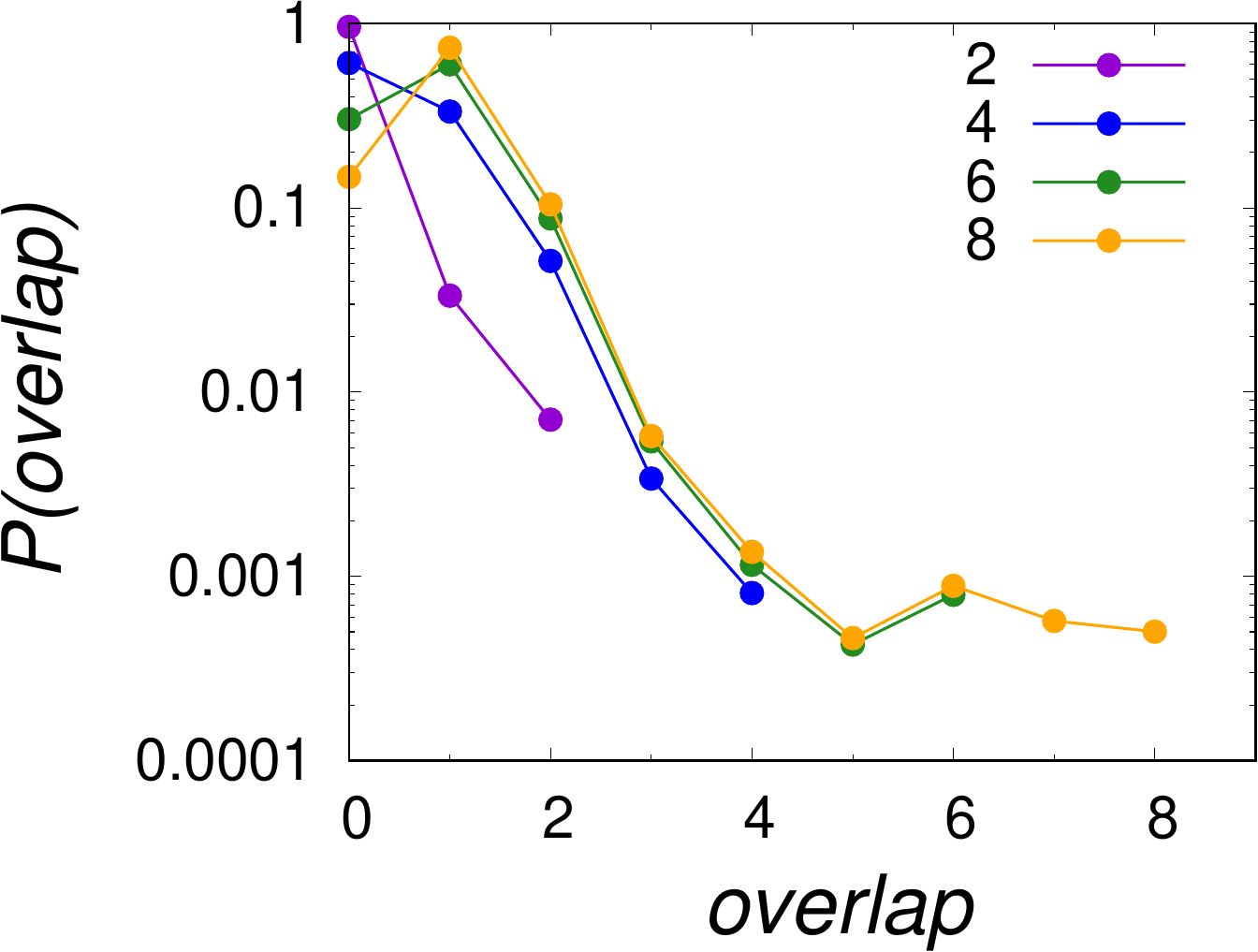}
\caption{Histograms of overlapping trajectory segments across all pairs of paths that reach each interface for the FFS calculation of XL hydrate nucleation. The value of the overlap indicates the number of configurations shared between a pair of paths.}
\label{fig:overlapHist-xl}
\end{figure}

While bottlenecks are detrimental to path diversity, it is reasonable to ask if early bottlenecks provide an opportunity for paths to diversify at later stages. We assess this by studying the overlap histograms (Fig.~\ref{fig:overlapHist-xl}). This histogram indicates that a  significant portion of the paths share one configuration. However, the probability of sharing $>$1 configuration decreases drastically for all interfaces. This suggests that the paths are indeed diversifying after the bottleneck, and indicates that we may be sufficiently sampling the transition path ensemble. To further investigate this aspect, analysis of the transition mechanisms and nucleation rates were performed on the entire path ensemble and the subset of the path ensemble comprised of trajectories that did not come from the one reactive configuration. We found that both ensembles gave us similar results. The details are described in Ref.~\onlinecite{DeFever:17:JCP}.

\begin{figure}[h]
\centering
\includegraphics[width=0.85\linewidth]{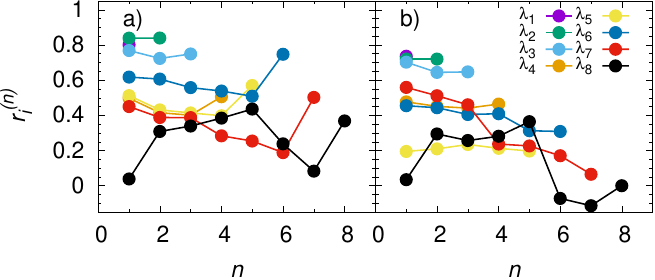}
\caption{Path correlation for the XL FFS paths at each interface calculated using the intraclass correlation of crossing probabilities. Each curve corresponds to an interface. In (a), all paths were considered. In (b), the highly reactive configuration at interface 0 and its descendants were excluded.}
\label{fig:pathcorrelation-xl}
\end{figure}

The final measure of the quality of the sampling is the uncertainty in the nucleation rate. As with the LJ FFS calculation, the uncertainty of the XL hydrates nucleation rate was estimated by calculating the path correlation. The results for $r_i^{(n)}$ are in Fig.~\ref{fig:pathcorrelation-xl}(a). For the LJ system, $r_i^{(n)}$ generally decreases monotonically. However, this is not the case for the XL system. Most notably, $r_i^{(n)}$ increases sharply for most of the interfaces as $n$ changes from $i-1$ to $i$. This is because all of the configurations from the one highly-reactive configuration at $i=0$ merge into a single group, and this group has a consistently higher crossing probability than the rest of the configurations at a given interface. By removing the highly reactive configuration at interface 0 and its descendants, the correlation generally decreases for all interfaces (Fig.~\ref{fig:pathcorrelation-xl}(b)). This also eliminates the sharp increase that occurs when $n=i$ for several of the interfaces. In the most conservative case, paths are only considered independent when they share no configurations, i.e., $L=i$ in Eq.~\eqref{eq:sigma}. The nucleation rate when using this criterion is $\log_{10}(J_{\text{FFS}}) = 32.1 \pm 5.6$ for a 95\% confidence interval. The units of $J_{\text{FFS}}$ are m$^{-3}$ s$^{-1}$. If the criterion for choosing $L$ changes such that $r_i^{(n)}$ cutoff values of $1/e$ and 0.5 are used, the uncertainty changes to 5.2 and 2.7, respectively due to the change in number of independent groups. This indicates that we have a reasonable estimate of the XL hydrate nucleation rate even with presence of the bottleneck. From 20 simulations with straightforward MD, the estimated nucleation rate was $\log_{10}(J_{\text{MD}}) = 31.8$,~\cite{DeFever:17:JCP} which agrees well with the FFS estimate.

The analysis of the XL FFS calculations demonstrates that the approaches proposed here can be used to identify bottlenecks on-the-fly. It is worthwhile to consider what can be modified when a bottleneck is discovered. For XL hydrates FFS, the bottleneck appeared at the initial interface. In such a case, a natural course of action would be to reevaluate the basin simulations. Analysis of the configurations at the initial interface using various CVs could provide insights into the underlying reason for the relatively high reactivity of the single configuration. In our case, the basin sampling did not appear problematic. However, calculation of the committor estimates of the configurations was informative---the bottleneck configuration was the only one at the first interface with a high committor value. 
This suggested that we needed either to change our simulation conditions such that nucleation is not accessible within the basin simulation timescales, or change our CV to be one that is a better approximation of the RC. The latter is often harder. In other cases, it is possible that the bottleneck is discovered at later interfaces.
Here as well, quick estimates of the committor could be informative. 
It is, however, difficult to prescribe a generic solution to resolving bottlenecks, as often the solutions are system dependent. 

\section{Conclusions}

Path sampling methods are becoming more accessible due to increasing computational power and availability of software for their implementation. Often these techniques are used to study systems with complex energy landscapes where ensuring that the sampling is correct becomes important and challenging. While there has been significant discussion on the theory of these methods, and some on the implementation and pitfalls, there is no step-by-step guide that can help "non-advanced" users in ensuring the correctness of their calculations. In this work, we bridge this gap. We focus on two popular path sampling methods---RETIS and FFS---using nucleation in different systems as examples. 

For RETIS, we discuss different types of well and ill behaved convergence of the sampling that can be encountered. Using nucleation in LJ systems as a demonstrative example, we provide diagnostics to identify the sources of poor convergence and suggest guidelines to tune parameters to improve the sampling quality. The primary tell of poor sampling convergence is found by observing the crossing histograms. Depending on their behavior, further diagnostic tools such as path correlation functions, shooting move acceptance probabilities, and shooting point diversity can be employed to help determine the source of convergence issues. Problems related to the poor choice of the sampling CV, $\lambda$, are also discussed. For FFS, we discuss methods to ensure proper basin sampling to obtain a large number of independent first-interface configurations as efficiently as possible. We also propose new heuristics to help determine the quality of sampling at each interface on-the-fly. Both ideal and non-ideal examples of TPE sampling are illustrated for RETIS and FFS.

Finally, while it may be tempting to compare the two methods, the resulting accuracy and performance are highly dependent on the optimality of the chosen parameters and algorithms. While searches for these optimal parameters may be feasible for highly-simplified models, such a search would be prohibitively expensive for anything more complex. RETIS and FFS also take advantage of the distribution of computing resources in very different ways---RETIS relies on generating new paths by modifying previous ones, restricting the parallelization to the number of interfaces running, while FFS can be scaled out to an arbitrarily high number of concurrent simulations, resources permitting. The trade-off is that FFS only generates the interface ensembles sequentially starting from the reactant basin, while RETIS samples the full range of ensembles simultaneously. Thus, the performance of each method relies on the system under study, the choice of algorithms and parameters, and the nature of the available computing resources. We therefore refrain from attempting such a comparison. Both RETIS and FFS have found a wide array of applications, and, regardless of the method used, care must be taken in setting up and assessing the sampling quality. We hope that the practical guidelines presented here will aid new and expert users to this end.

\begin{acknowledgments}
J.R. acknowledges financial support from the Deutsche Forschungsgemeinschaft (DFG) through the Heisenberg Programme project 428315600. This material is based upon work supported by the U.S. Department of Energy, Office of Science, Office of Basic Energy Sciences, under Award Number DE-SC0015448. S.S. acknowledges the National Science Foundation (CAREER grant award No. 1653352) and R{\"u}hr University RESOLV center for support to travel to and stay in Germany. Clemson University is acknowledged for generous allotment of compute time on the Palmetto cluster. G.D.L. acknowledges support from Conacyt-Mexico through fellowship Ref. No. 220644 and from the Isaac Newton Trust Grant Ref No: 20.40(h).
\end{acknowledgments}

\section*{Data Availability Statement}
The data that support the findings of this study are available from the corresponding author upon reasonable request. The data will also be made available in the near future at https://github.com/sarupriagroup.

\appendix*
% remove * if more than one appendix

\section{Intraclass correlation for FFS}

To calculate the ICC for FFS interfaces, the notion of groups of configurations at each interface described in Sec.~\ref{sec:ffs-lj} is used. The set of groups for each $i$ and $n$ combination is denoted as $\mathcal{G}_i^{(n)}$. $|\mathcal{G}_i^{(n)}|$ is the number of groups at interface $i$ based on ancestry at $i-n$. The number of independent groups at each interface is determined by calculating the ICC ($r_i^{(n)}$) for each $(i,n)$ pair. The ICC across configurations is calculated using the crossing probability of each configuration:
\begin{equation}
    r_i^{(n)} = \frac{MSB - MSW}{MSB + (N_G-1)MSW}
\end{equation}
where, $MSB$ is the between-group mean square, $MSW$ is the within-group mean square, and $N_G$ is the group size. These are calculated as follows for each $(i,n)$ pair:
\begin{equation}
    MSB = \frac{1}{|\mathcal{G}_i^{(n)}|(|\mathcal{G}_i^{(n)}|-1)} \sum_{g \in \mathcal{G}_i^{(n)}}{|g|(p_{i,g}-p_i)^2}
\end{equation}
\begin{equation}
    MSW = \frac{1}{(N_i-|\mathcal{G}_i^{(n)}|)} \sum_{g \in \mathcal{G}_i^{(n)}} \sum_{j \in g}{(p_{i,j}-p_{i,g})^2}
\end{equation}
\begin{equation}
    N_G = \frac{N_i^2 - \sum_{g \in \mathcal{G}_i^{(n)}}{|g|^2}}{(|\mathcal{G}_i^{(n)}|-1)N_i}
\end{equation}

$N_i$ is the number of configurations at interface $i$, $p_i$ is equivalent to $P_{A}(\lambda_{i+1}|\lambda_i)$, and $p_{i,j}$ is $P_{A}(\lambda_{i+1}|\lambda_i)$ of configuration $j$ at interface $i$. $p_{i,g}$ is the crossing probability averaged over configurations that belong to group $g$ and $|g|$ is the number of configurations in group $g$. For each $i$, the smallest value of $n$ is selected such that $r_i^{(n)} < c$, where $c$ is a cutoff value that can be modified. The chosen value of $n$, denoted $L$, is then used to form groups of configurations at each $i$. These groups are then used to calculate the standard error of the mean of $p_i$, $\sigma_{p_i}$:
\begin{equation}
    \label{eq:sigma}
    \sigma_{p_{i}}=\sqrt{\frac{1}{\left|\mathcal{G}_{i}^{(L)}\right|\left(\left|\mathcal{G}_{i}^{(L)}\right|-1\right)} \sum_{g \in \mathcal{G}_{i}^{(L)}}\left(p_{i}^{(L)}-p_{i,g}\right)^{2}}
\end{equation}
$p_i^{(L)}$ is the average crossing probability for $n=L$ when each group of configurations is treated as a single configuration. For a given value of $n$, $p_{i}^{(n)}$ is calculated with the following:
\begin{equation}
    p_{i}^{(n)}=\frac{1}{\left|\mathcal{G}_{i}^{(n)}\right|} \sum_{g \in \mathcal{G}_{i}^{(n)}} \frac{1}{|g|} \sum_{j \in g} p_{i, j}
\end{equation}

\end{document}